\documentclass[fleqn,usenatbib]{mnras}

\usepackage{newtxtext,newtxmath}

\usepackage[T1]{fontenc}
\usepackage{ae,aecompl}
\usepackage{bm}

\usepackage{graphicx}	
\usepackage{amsmath}	
\usepackage{amssymb}	

\title[The $E_\text{p,i}$ -- $E_\text{iso}$ correlation]{The $\bm{E}_\text{p,i}$ -- $\bm{E}_\text{iso}$ correlation: \\ type I gamma-ray bursts and the new classification method}

\author[P. Y. Minaev et al.]{
P. Y. Minaev,$^{1,2}$\thanks{E-mail: minaevp@mail.ru}
A. S. Pozanenko,$^{1,2,3}$
\\
$^{1}$ Space Research Institute of the Russian Academy of Sciences (IKI), 84/32 Profsoyuznaya Str., Moscow, 117997, Russia \\
$^{2}$ Moscow Institute of Physics and Technology (MIPT), 9 Institutskiy per., Dolgoprudny, 141701, Russia \\
$^{3}$ National Research University Higher School of Economics, Moscow, 101000, Russia
}

\date{Accepted XXX. Received YYY; in original form ZZZ}

\pubyear{2019}

\begin{document}
\label{firstpage}
\pagerange{\pageref{firstpage}--\pageref{lastpage}}
\maketitle

\begin{abstract}
We present the most extensive sample of 45 type I (short) and 275 type II (long) gamma-ray bursts (GRB) with known redshift to investigate the correlation between the rest frame peak energy, $E_\text{p,i}$ and the total isotropic equivalent energy, $E_\text{iso}$ of the prompt emission (Amati relation). The $E_\text{p,i}$ -- $E_\text{iso}$ correlation for type I bursts is found to be well-distinguished from the one constructed for type II bursts and has a similar power-law index value, $E_\text{p,i}$ $\sim$ $E_\text{iso}^{~0.4}$, which possibly indicates the same emission mechanism of  both GRB types. We show that the initial pulse complex (IPC) of type I bursts with an extended emission and regular type I bursts follow the same correlation. We obtain similar results for type II bursts associated with Ic supernovae and for regular type II bursts. Three possible outliers from the $E_\text{p,i}$ -- $E_\text{iso}$ correlation for type II subsample are detected. Significant evolution of the $E_\text{p,i}$ -- $E_\text{iso}$ correlation with redshift for type II bursts is not found. We suggest the new classification method, based on the $E_\text{p,i}$ -- $E_\text{iso}$ correlation and introduce two parameters, $EH = E_\text{p,i,2} ~ E_\text{iso,51}^{~-0.4}$ and $EHD = E_\text{p,i,2} ~ E_\text{iso,51}^{~-0.4} ~ T_\text{90,i}^{~-0.5}$, where $E_\text{p,i,2}$ is the value of $E_\text{p,i}$ parameter in units of 100 keV, $E_\text{iso,51}$ is the value of $E_\text{iso}$ parameter in units of $10^{51}$ erg and $T_\text{90,i}$ is the rest frame duration in units of seconds. $EHD$ is found to be the most reliable parameter for the blind type I / type II classification, which can be used to classify GRBs with no redshift.
\end{abstract}




\section{Introduction}
\label{sec:intro}

Two classes of gamma-ray bursts (GRB) were discovered in a series of KONUS experiments forty years ago \citep{maz81} and later confirmed in the Phebus/GRANAT experiment \citep{deza92}. The classification scheme based on the bimodal nature of the duration distribution with the separation at $T_\text{90}$ $\sim$ 2 s was proposed in the BATSE/CGRO experiment, where $T_\text{90}$ parameter (time interval with integrated counts raise from 5\% to 95\%) was introduced \citep{kou93}. Type I (short) bursts were also found to have a harder energy spectrum (e.g., in terms of a hardness ratio -- the ratio of energy fluxes in different energy bands) and less distinct spectral evolution (lag) comparing with type II (long) ones \citep{kou93,nor05,min10a,min10b,min12,min14}.

$T_\text{90}$ and hardness ratio distributions, used for the classification of GRBs traditionally, are overlapped significantly. It makes the problem of correct blind classification still very actual, especially in the new era of gravitational wave astronomy. The one and only confirmed gamma-ray burst for the moment, associated with gravitational waves, GRB 170817, is placed in the middle of the overlap region \citep{abb17,abb17b,gol17,poz18a}.

Moreover, $T_\text{90}$ parameter is highly affected by the number of selection effects. Firstly, it depends on an energy range used for the calculation: a pulse narrowing with an energy follows a power-law with the index of $\sim$ -0.4 \citep[e.g.][]{fen95,min10b}. The energy range used for $T_\text{90}$ calculation varies with experiments: (15, 150) keV range is typical for Swift, (10, 1000) keV -- for Fermi, and (80, 1200) keV -- for Konus-Wind. The pulse narrowing with the energy leads to shifting of $T_\text{90}$ separation line, used for a blind classification, with energy range \citep{min10b}. Additionally, $T_\text{90}$ depends on the detector sensitivity and, in particular, on background variations (brighter events with a stable background are longer). Next, the observed duration depends on redshift of the source $T_\text{90}$ = $T_\text{90,i}(1+z)$, where $T_\text{90,i}$ is the intrinsic (rest-frame) duration. Finally, the energy range used for $T_\text{90}$ calculation is shifted to lower energies from the one in the rest frame. All these effects make an estimation of the ``true'' (not biased) $T_\text{90,i}$ value virtually impossible. Therefore, $T_\text{90,i}$ values should be used in the analysis with caution, meaning the classification of GRBs based only on their duration is unreliable.

Apart from the duration and spectral hardness, the spectral lag, a parameter characterizing the spectral evolution, was used  to classify GRBs: type I bursts have negligible lags while larger positive spectral lags are characteristic of type II GRBs \citep[e.g.][]{yi06, norbon06, zha06a, min12,min14}. An empirical correlation of the lag on luminosity has been detected for the class of type II GRBs \citep[e.g.][]{nor00, nor02, geh06, sch07, hakk08, ukw12, bern15}. \citet{hakk11,min14} showed that the observed properties of short pulses in type II GRBs with a complex multipulse structure are similar to those of type I GRBs. \citet{min14} also revealed that the lag value is strongly affected by superposition effects in multipulsed bursts: one can obtain negligible lag even for long type II GRB, if it is characterized by a complex structure of light curve. Therefore, at this stage we would ignore the spectral lags as classification indicators. More parameters useful for classification are also discussed e.g. in \citet{don06,ber14}. One more evident criterion can be added, the detection of kilonova emission for type I bursts \citep[e.g.][]{jin16,yang15,tan16,poz18a,pan19}.

The correct classification is crucial for understanding GRB progenitors. Type I bursts are associated with a merger of compact binaries, consisted of two neutron stars \citep{blin84,pac86,mes92,ros03}, which was recently confirmed by the gravitational wave experiments LIGO-Virgo for GRB 170817A \citep{abb17,abb17b,gol17,sav17,poz18a}. Several type I bursts are also accompanied by an additional component with a duration of tens of seconds and a soft energy spectrum -- the extended emission (EE), which was detected both in the light curves of several individual events and in the averaged light curve of a group of events \citep{laz01,con02,mon05,geh06,min10a,min10b,nor10}. The nature of the extended emission is still unclear, it could be related to an activity of a magnetar formed in a merger process \citep{met08}, beginning of an afterglow \citep{min10a}, fallback accretion \citep{ross07,bar11,gibs17}. The extended emission sufficiently complicates the GRB classification, confirming that the classification based on the only $T_\text{90}$ value is not robust.

Type II bursts are associated with a core collapse of supermassive stars (e.g. \citet{woo93,pac98,mes06}), $\sim$ 10 \% of type II GRBs with an identified optical component and a measured redshift are also accompanied by a bright (intrinsically) Ic supernovae (e.g. \citet{gal98,hjo03,pac98,kul98,can17,vol17}). It is unclear, whether the supernova (SN) is the common feature of all type II bursts and its non-detection is connected with only selection effects for the large sample of bursts (supernovae are harder to detect at higher redshift).

Gamma-ray bursts are characterized by a number of correlations between different observational parameters. One can mention hardness -- peak flux ($E_\text{p}$ -- $f_\text{p}$) and hardness -- fluence ($E_\text{p}$ -- $F_\text{tot}$) correlations (e.g. \citet{mal95, dez97, lloyd00}), which were found in the epoch, when distances to GRB sources were not known. The hardness indicator, spectral parameter $E_\text{p}$ is the position of the extremum (maximum) of the energy spectrum $\nu F_{\nu}$. Later, when the era of optical observations with successful redshift measuring began, these correlations were transformed to intrinsic hardness -- peak luminosity ($E_\text{p,i}$ -- $L_\text{iso}$) and intrinsic hardness -- total energy ($E_\text{p,i}$ -- $E_\text{iso}$) correlations, also known as Yonetoku and Amati relations, respectively (e.g. \citet{yon04, ama02}). $E_\text{iso}$ ($L_\text{iso}$) is the isotropic equivalent total energy (peak luminosity), emitted in gamma-rays in (1, 10000) keV range, $E_\text{p,i}$ is the intrinsic peak energy, $E_\text{p,i}$ = $E_\text{p}(1+z)$. Observed in many cases, break in afterglow light curves confirmed the jet-like non-isotropic emission behavior of GRBs and the necessity of $E_\text{iso}$ correction to estimate the collimated energy release $E_{\gamma}$ = $E_\text{iso}(1-\cos\theta_\text{jet})$ and collimated luminosity $L_{\gamma}$ = $L_\text{iso}(1-\cos\theta_\text{jet})$, where $\theta_\text{jet}$ is the jet opening angle, derived from the time of the afterglow jet break. The hardness -- collimation corrected energy ($E_\text{p,i}$ -- $E_{\gamma}$) correlation was also found, known as Ghirlanda relation \citep[e.g.][]{ghir04}.

We investigate the $E_\text{p,i}$ -- $E_\text{iso}$ correlation (Amati relation). Previously it was analyzed mostly for type II bursts \citep[e.g.][]{ama02, ama06, tsv17}. Type I bursts were found to be outliers of the $E_\text{p,i}$ -- $E_\text{iso}$ correlation for type II bursts, so the correlation could also be used for the classification of GRBs \citep[e.g.][]{lu10,zha12,qin13, zha18a, shah15, zou18, poz18a}. More over, in \citet{poz19b} the Amati relation was used to estimate the $E_\text{p,i}$ value for the second LIGO/Virgo event S190425z connected with binary neutron star merger, which was observed in gamma-ray domain by SPI-ACS/INTEGRAL experiment. Hardness -- fluence ($E_\text{p}$ -- $F_\text{tot}$) correlation, constructed for GRBs with no redshift, also demonstrated bimodal nature \citep[e.g.][]{ghir09}. The sample of investigated type I bursts with measured redshift $E_\text{p,i}$ and $E_\text{iso}$ parameters was limited in most previous works by $\sim$ 20 events or even less \citep[e.g.][]{lu10,qin13,tsv17,poz18a}, making the behavior of the correlation for type I bursts to be unresolved. The sample of type I bursts in \citet{zha18a} reached 31 events and indicated the same value of power-law index of the Amati relation for both type I and type II classes.

The nature of the $E_\text{p,i}$ -- $E_\text{iso}$ correlation is unknown. In the trivial case, it could be connected with strong selection effects. One of the probable explanations considers viewing angle effects: the smaller the angle between the observer's line of sight and the jet axis is, the brighter and harder the gamma-ray emission will be. Under the assumption, the correlation is $E_\text{p,i} \sim E_\text{iso}^{~1/3}$ if it is a cone relativistic jet emission \citep[e.g.][]{eic04,lev05,poz18a}. So the observable behavior of the correlation could reveal some features of an emission mechanism and a structure of the jet, which is substantial when comparing two types of GRB progenitors. Recently it was shown that the Yonetoku relation (the $E_\text{p,i}$ -- $L_\text{iso}$ correlation) also can be explained by the viewing angle effects \citep{ito19}.

In the paper, we construct the sample of 45 type I and 275 type II gamma-ray bursts with known redshift and well constrained $E_\text{p}$ and analyze its features, including various selection effects (Section~\ref{sec:sample}). We investigate the $E_\text{p,i}$ -- $E_\text{iso}$ correlation for different subsamples (Section~\ref{sec:corr}) and suggest the new classification method of GRBs, introducing two parameters ($EH = E_\text{p,i,2} ~ E_\text{iso,51}^{~-0.4}$ and $EHD = E_\text{p,i,2} ~ E_\text{iso,51}^{~-0.4} ~ T_\text{90,i}^{~-0.5}$), to make the blind classification of the bursts more robust (Section~\ref{sec:corr_class}).

\begin{table}

 \caption{The number of GRBs, depending on experiments used for $E_\text{iso}$, $E_\text{p,i}$ and $T_\text{90,i}$ calculation.}  \label{ta_exp_stat}

 \begin{tabular}{lcccc}
  \hline

Experiment	&	Type I &	Type II	&	Total & $R^a$	\\	\hline
Konus-Wind	&	12	&	157	&	169	& 0.07 \\	
BeppoSAX	&	0	&	4	&	4	& 0 \\	
BATSE/CGRO	&	0	&	2	&	2	& 0 \\	
HETE-2	&	1	&	11	&	12	&  0.08 \\	
Swift	&	17	&	67	&	84	&  0.20 \\	
Fermi	&	15	&	34	&	49	&  0.31 \\	\hline
Total	&	45	&	275	&	320	&  0.14 \\	\hline

\multicolumn{5}{l}{$^a$ - $R$ = Type I / Total }

 \end{tabular}
\end{table}

\section{The sample}
\label{sec:sample}

\subsection{The sample construction}
\label{sec:sample_constr}

\begin{table*}


 \caption{Observed parameters of investigated subsamples.}  \label{ta_par}

 \begin{tabular}{lcccccccc}
  \hline

Subsample	&	N	&	$z$ [min]	&	$z$ [median]	&	$z$ [max]	&	$E_\text{iso}$ [median]	&	$E_\text{p,i}$ [median]	&	$T_\text{90,i}$ [median]	\\	
	&		&		&		&		&	$10^{51}$ erg	&	keV	&	s	\\	\hline
I	&	45	&	0.0097	&	0.46	&	3.05	&	0.68	&	706	&	0.27	\\	
I + EE	&	11	&	0.125	&	0.41	&	1.72	&	3.2	&	909	&	0.58	\\	
I w/o EE	&	34	&	0.0097	&	0.51	&	3.05	&	0.57	&	662	&	0.25	\\	\hline
II	&	275	&	0.0085	&	1.60	&	8.10	&	100	&	446	&	14.5	\\	
II + SN	&	40	&	0.0085	&	0.55	&	1.06	&	8.98	&	134	&	23.8	\\	
II w/o SN	&	235	&	0.12	&	1.90	&	8.10	&	122	&	512	&	13.6	\\	\hline
II $z_1$	&	55	&	0.0085	&	0.53	&	0.76	&	9.15	&	218	&	24.6	\\	
II $z_2$	&	55	&	0.78	&	1.00	&	1.31	&	70.4	&	372	&	13.1	\\	
II $z_3$	&	55	&	1.32	&	1.60	&	2.01	&	130	&	540	&	14.5	\\	
II $z_4$	&	55	&	2.01	&	2.30	&	2.82	&	230	&	694	&	10.0	\\	
II $z_5$	&	55	&	2.82	&	3.65	&	8.10	&	176	&	575	&	11.2	\\	\hline

 \end{tabular}
\end{table*}

To construct the sample of gamma-ray bursts with a measured redshift (both spectroscopic and photometric) and $E_\text{p}$ parameters we used different sources:\\
1. GRB catalogs of Konus/Wind \citep{svi16,tsv17}, BeppoSAX \citep{fro09}, GBM/Fermi \citep{bha16} experiments; \\
2. Papers concerning $E_\text{p,i}$ -- $E_\text{iso}$ correlation investigation (e.g. \citet{ama06,qin13,zha18a,zou18});\\
3. Papers concerning analysis of individual GRBs (e.g. \citet{min17,poz18a,pan19}); \\
4. GCN circulars archive \footnote{\url{https://gcn.gsfc.nasa.gov/gcn3_archive.html}}; \\
5. Online catalog of well-localized GRBs \footnote{\url{http://www.mpe.mpg.de/~jcg/grbgen.html}}; \\
6. The catalog of GRB associated supernova \citep{can17}; \\
7. The paper related to the extended emission investigation of type I GRBs \citep{nor10}.

We also derive $E_\text{iso}$ and $E_\text{p,i}$ values for 45 GRBs using spectral parameters and redshift, available in the literature. The $E_\text{iso}$ parameter is calculated via $E_\text{iso} = \frac{4\pi D_\text{L}^{~2} F}{1+z}$, where $F$ is the burst fluence in (1, 10000) keV energy range in the rest frame and $D_\text{L}$ is the luminosity distance calculated using standard cosmological parameters, $H_0$ = 67.3 \,km\,s$^{-1}$\,Mpc$^{-1}$, $\Omega_{\rm M}$ = 0.315, $\Omega_\Lambda$ = 0.685 \citep{planck14}. $E_\text{p,i}$ parameter is calculated via $E_{\rm p,i} = E_{\rm p}(1+z)$. The duration of a burst in the rest frame is calculated as $T_\text{90,i}$ = $T_\text{90}/(1+z)$.

The complete sample of GRBs including 45 type I and 275 type II bursts registered up to 2019, January is presented in Appendix, Table~\ref{ta1}. 3 type I bursts and 13 type II bursts have photometric redshift estimation. 11 type I bursts are characterized by an extended emission (EE) component, 40 type II bursts are associated with Ic supernova (19 photometrical and 21 spectroscopical ones). For type I+EE bursts we use $T_\text{90,i}$, $E_\text{p,i}$ and $E_\text{iso}$ values obtained for the initial pulse complex (IPC) only, ignoring an extended emission component, probably having a different nature. We do not perform any classification analysis at this step and use the classification results from the literature, usually obtained from the complex of different critera.

The Table~\ref{ta_exp_stat} summarizes the statistics of the experiments used for calculation $E_\text{iso}$ and $E_\text{p,i}$ values. $R$ parameter is defined as the ratio Type I / Total and indicates the utility of the experiment for a classification. The Konus-Wind experiment has been operated since 1994, has a good design (wide energy range, crucial for $E_\text{p,i}$ determination, in particular) and, as a consequence, it is the ``leading'' experiment in our sample (169 GRBs, $R$ = 0.07). To maximize the homogeneity of the sample we use the Konus-Wind derived observational parameters in case of GRBs, registered by several experiments (e.g., also by Fermi or Swift). Swift (84 GRBs, $R$ = 0.20) is a perfect instrument for a fast and precise localization of GRBs, which is crucial for optical observations and redshift determination; but its capability to measure the $E_\text{p}$ value is  limited due to the soft and narrow energy range of the BAT instrument. Fermi (49 GRBs, $R$ = 0.31) can not provide a precise localization, but it is excellent for the precise $E_\text{p,i}$ determination, complementing the Swift.

\subsection{Features of the sample}
\label{sec:sample_prop}

The dependence of $T_\text{90,i}$, $E_\text{iso}$ and $E_\text{p,i}$ parameters on redshift for both GRB types and their subsamples is presented at Fig.~\ref{fig:z}. The corresponding statistics is summarized in  Table~\ref{ta_par}. We also form five equal subsamples of type II bursts (55 bursts each), based on $z$ values: the $z_1$ subsample represents the closest bursts while the $z_5$ one - the most distant bursts. We use them to check the evolution of $E_\text{p,i}$ -- $E_\text{iso}$ correlation with redshift (Section~\ref{sec:corr_evo}).

\subsubsection{Redshift, $z$}
\label{sec:sample_prop_z}

Type I bursts are significantly closer than type II ones: the median values are $z$ = 0.46 and $z$ = 1.60, respectively. Firstly, it should be connected with selection effects. In general, optical afterglows (crucial for redshift determination) of type I bursts are fainter than ones of type II bursts, and therefore are harder to detect at high redshift \citep[e.g.][]{kann11}. Secondly, type I bursts are generated by a merger of compact components in old binary systems, which could not occur at a high redshift due to a long evolution of the system towards merging \citep[e.g.][]{gue05}. In our sample, the most distant type I burst is GRB 081024B with $z$ = 3.05, the most distant type II burst is GRB 090423 with $z$ = 8.1 (both values are not spectroscopic).

Type I bursts with an extended emission (I+EE) do not differ from the regular ones in terms of redshift distribution, having similar $z$ median values. It reveals negligible selection effects of the extended emission detection: if the extended emission is a weak component, the redshift distribution for I+EE bursts is expected to be shifted towards low redshift values. It indicates the same progenitor type for both subsamples (e.g. no difference in a merging delay), but the presence of two separate subclasses of type I bursts.

Type II bursts associated with Ic supernova (II+SN) demonstrate the opposite behavior. The redshift distribution for II+SN bursts is significantly shifted to low values comparing with regular type II bursts (median values are $z$ = 0.55 and $z$ = 1.90, correspondingly). The difference is obviously connected with strong selection effects: supernova signatures (photometric, and spectroscopic especially) are very difficult to reveal at large distances ($z \ga 1$) with present optical telescopes. It is confirmed by the fact that the closest type II bursts ($z \la $ 0.3) are mostly accompanied by supernovae (see Fig.~\ref{fig:z}), which may indicate that all type II GRBs are accompanied by Ic supernovae. The biased redshift distribution for II+SN bursts leads to significant biases in other parameter distributions (Sections~\ref{sec:sample_prop_dur},~\ref{sec:sample_prop_eiso},~\ref{sec:sample_epi}).

\subsubsection{Duration, $T_\text{90,i}$}
\label{sec:sample_prop_dur}

\begin{figure*}
	\includegraphics[width=2.0\columnwidth]{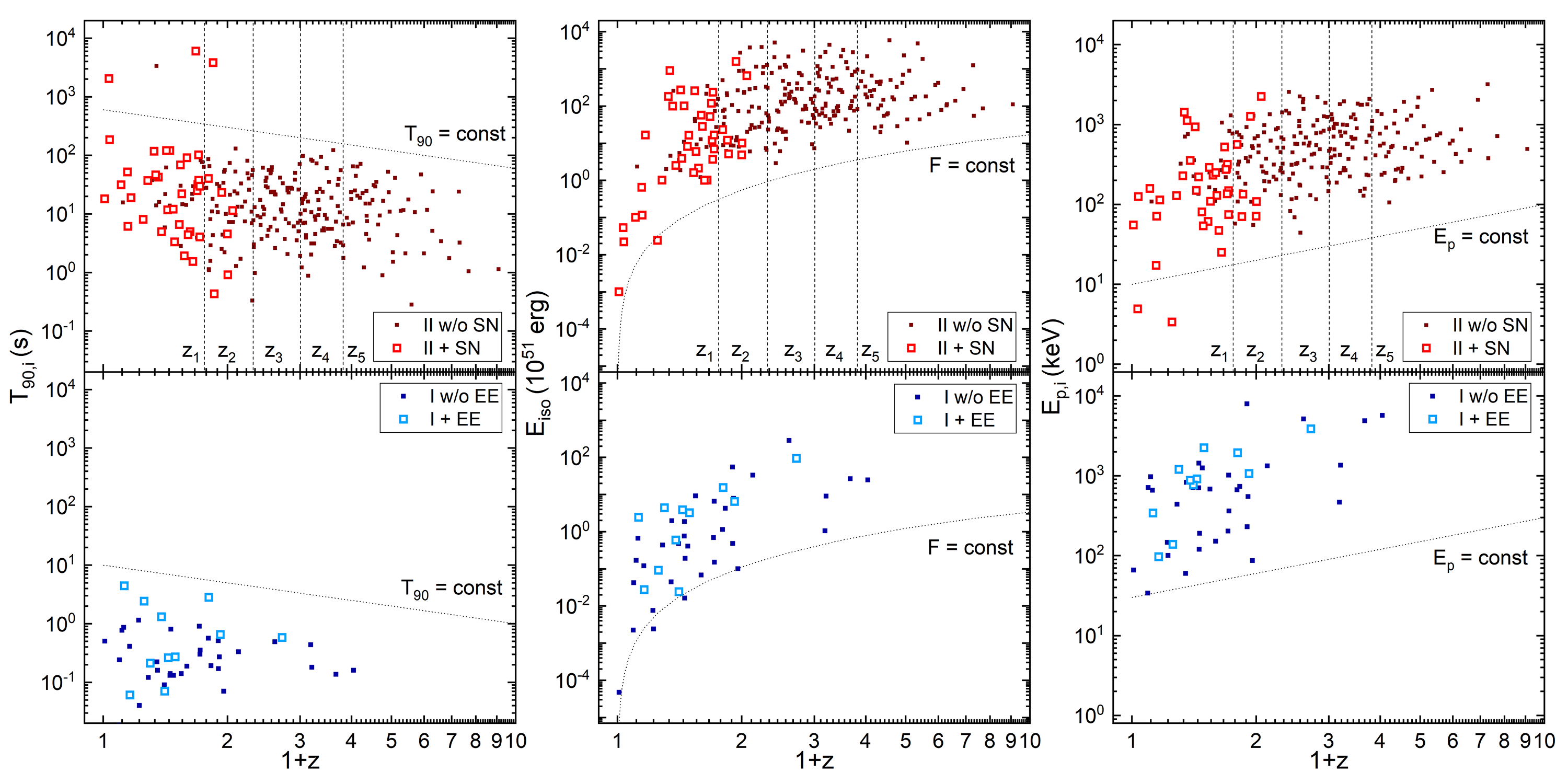}
    \caption{The dependence of $T_\text{90,i}$ (left), $E_\text{iso}$ (middle) and $E_\text{p,i}$ (right) parameters on redshift, for regular type I bursts (blue dots) and type I bursts with extended emission (cyan squares) at bottom figures, regular type II bursts (dark red dots) and type II bursts associated with Ic supernova (red squares) at top figures. Vertical dashed lines show bounds of $z_1$ -- $z_5$ subsamples of type II bursts. Dotted lines represent the dependence of intrinsic rest frame parameters on redshift for parameters fixed in the observer frame.}
    \label{fig:z}
\end{figure*}

The dependence of $T_\text{90,i}$ on redshift for both GRB types is presented at the left graphs of Fig.~\ref{fig:z}.

Type I bursts are 50 times shorter than type II ones: the median values are $T_\text{90,i}$ = 0.27 s and $T_\text{90,i}$ = 14.5 s, correspondingly. Four type I bursts are longer 1 s and four type II bursts are shorter 1 s, confirming high overlap of a duration distribution and its weak reliability as a blind classification method.

Our sample contains four ultralong bursts with $T_\text{90,i}$ > 1000 s. We find no peculiarities of these bursts except the duration. The nature of ultralong GRBs and their possible separation into a different burst class are still debatable \citep[e.g.][]{gen19}.

The lack of bursts with a duration in the interval (200, 2000) s could be connected (at least, partially) with selection effects. Long GRBs ($T_\text{90}$ > 200 s) are difficult to register due to background variations and due to source occultation by the Earth in experiments on low orbits (Fermi, Swift) or due to telemetry limitations (Konus-Wind). Therefore, the $T_\text{90,i}$ distribution becomes asymmetric for type II bursts, having sharper ``long'' edge and well fitted by a skewed gaussian (see Section~\ref{sec:corr_class}, equation~(\ref{eq:gauss}) and Fig.~\ref{fig:class}). In addition, the rest frame duration, $T_\text{90,i} = \frac{T_\text{90}}{1+z}$ decreases with redshift (shown by dotted lines at left graphs in Fig.~\ref{fig:z}). As a consequence, the duration of long and distant bursts is underestimated most probably due to selection effects (see the $z_5$ subsample of type II bursts at Fig.~\ref{fig:z}).

There is also a lack of relatively short ($T_\text{90,i}$ < 4 s) type II bursts at low redshift ($z$ < 0.4, $z_1$ subsample). These bursts are also fainter ($E_\text{iso}$ < 10$^{52}$ erg) and softer ($E_\text{p,i}$ < 200 keV) comparing with more distant bursts (see Fig.~\ref{fig:z}). In \citet{nor05} was shown, that long bursts with wide pulses are also fainter and softer. Therefore, the lack is possibly connected with selection effects.

All discussed selection effects can lead to a negative $z$ -- $T_\text{90,i}$ correlation for type II bursts. We check the possible correlation using the Spearman rank-order correlation coefficient ($\rho$) and the associated null-hypothesis (chance) probability, or $P_{\rho}$ value. The null hypothesis is that no correlation exists; therefore, a small $P_{\rho}$ value indicates a significant correlation. We apply this method to analyze other correlations through the paper. As expected, we do not detect the correlation for type I bursts ($\rho$ = -0.006, $P_{\rho}$ = 0.97), but find the negative correlation for type II bursts ($\rho$ = -0.21, $P_{\rho}$ = 5.2 $\times~ 10^{-4}$). The result is also confirmed by a slight decreasing of median $T_\text{90,i}$ values with increasing of $z$ for $z_1$ -- $z_5$ type II bursts subsamples (see Table~\ref{ta_par}).

I+EE bursts are longer than regular type I bursts: median values are $T_\text{90,i}$ = 0.58 s and $T_\text{90,i}$ = 0.25 s, correspondingly. Most (four of five) type I bursts, having $T_\text{90,i}$ > 1 s, are from type I+EE subsample. It possibly confirms the definition of I+EE bursts as a separate subclass of type I bursts. As one of the possible scenarios, a merging of neutron stars under some circumstances (affecting also prompt emission properties) could result in the forming of magnetar, responsible for an extended emission component \citep[e.g.][]{met08}.

As was shown earlier, a subsample of II+SN bursts is affected by strong selection effects. II+SN bursts, representing mostly the closest $z_1$ subsample, are longer than regular type II bursts, as a consequence: median values are $T_\text{90,i}$ = 23.8 s and $T_\text{90,i}$ = 13.6 s, correspondingly. Nevertheless, they cover a whole range of $T_\text{90,i}$ values for type II bursts, confirming the suggestion, that they do not differ from regular type II bursts, meaning all type II bursts are accompanied by SN Ic.

\subsubsection{Total isotropic equivalent energy release, $E_\text{iso}$}
\label{sec:sample_prop_eiso}

Type I bursts are 100 times fainter than type II bursts: median values are $E_\text{iso}$ = 6.8 $\times~ 10^{50}$ erg and $E_\text{iso}$ = 1.0 $\times~ 10^{53}$ erg, correspondingly.

The dependence of $E_\text{iso}$ on redshift for both GRB types is presented in the middle graphs of Fig.~\ref{fig:z}. The dependence is highly affected by selection effects for both GRB types. Dotted curves at Fig.~\ref{fig:z} show the $E_\text{iso}(z)$ dependence for a fixed fluence $F$, representing a sensitivity of a gamma-ray detector to triggering a burst. As a consequence, we can detect weak bursts only at low redshift: the bottom-right side of the $z$ -- $E_\text{iso}$ diagram is empty due to selection effects.

One can see also at Fig.~\ref{fig:z} the increasing of the number of bright bursts with increasing of a redshift for both GRB types. It can be explained by the non-flat luminosity function for GRBs: bright bursts are more rare, than weak ones.

All these selection effects lead to a significant positive $z$ -- $E_\text{iso}$ correlation for both GRB types: $\rho$ = 0.69 and $P_{\rho}$ = 1.6 $\times~ 10^{-7}$ for type I bursts, $\rho$ = 0.46 and $P_{\rho}$ = 1.9 $\times~ 10^{-15}$ for type II bursts. It is also supported by median values of $z_1$ -- $z_5$ subsamples (see Table~\ref{ta_par}).

I+EE bursts are brighter than regular ones in terms of median values ($E_\text{iso}$ = 3.2 $\times~ 10^{51}$ erg and $E_\text{iso}$ = 0.57 $\times~ 10^{51}$ erg, correspondingly), but they are not concentrated to the top of the $E_\text{iso}$ distribution and cover a whole range of observable $E_\text{iso}$ values of type I bursts. Therefore, we cannot draw a definitive conclusion about their difference in $E_\text{iso}$ from regular type I bursts. It could be accidental.

II+SN bursts are significantly fainter than regular type II bursts in terms of median values: $E_\text{iso}$ = 9.0 $\times~ 10^{51}$ erg and $E_\text{iso}$ = 1.22 $\times~ 10^{53}$ erg, correspondingly. It could be explained by the selection effects, described above (Section~\ref{sec:sample_prop_z}). The covering of the whole range of $E_\text{iso}$ distribution by them supports the suggestion.

\subsubsection{The position of the extremum (maximum) in $\nu F_{\nu}$ spectrum, $E_\text{p,i}$}
\label{sec:sample_epi}

The dependence of $E_\text{p,i}$ on redshift for both GRB types is presented in the right graphs of Fig.~\ref{fig:z}.

Type I bursts are 1.5 times harder ($E_\text{p,i}$ = 706 keV vs $E_\text{p,i}$ = 446 keV), than type II bursts. The behavior of all subsamples on $z$ -- $E_\text{p,i}$ diagram is analogous to their behavior on $z$ -- $E_\text{iso}$ diagram. It could be explained by the presence of the $E_\text{p,i}$ -- $E_\text{iso}$ correlation. We find significant positive $z$ -- $E_\text{p,i}$ correlation for type II bursts ($\rho$ = 0.40 and $P_{\rho}$ = 3.7 $\times~ 10^{-12}$) and the weak positive correlation for type I bursts ($\rho$ = 0.52 and $P_{\rho}$ = 2.4 $\times~ 10^{-4}$).

A possible selection effect at the $z$ -- $E_\text{p,i}$ diagram is connected with a limited working energy range of gamma-ray detectors. If $E_\text{p,i}$ is placed outside the range, it could not be determined. A typical value of a lower energy range bound of modern detectors is $\sim$ 10 keV. The corresponding trajectory $E_\text{p,i} = E_\text{p}(1+z)$ is shown by the dotted curve at top right graph in Fig.~\ref{fig:z}. At high redshift ($z$ = 5) it becomes $E_\text{p,i}$ = 60 keV, while the minimal observed value is $\sim$ 5 times larger (type II bursts sample). It means the possible selection effect of the detector lower energy range bound is minimal.

The situation with the upper energy bound is more complicated, because the $E_\text{p,i}$ determination depends also on count statistics, which is low at high energies due to the hard intrinsic GRB energy spectrum and due to decreasing of the effective area of detectors. The influence of this effect needs detailed investigation, which is beyond the scope of the paper. Here, we suppose this effect to be small and consider the evolution of $E_\text{p,i}$ with redshift to be real, suggesting the connection of observable $z$ -- $E_\text{p,i}$ correlation with $z$ -- $E_\text{iso}$ and $E_\text{p,i}$ -- $E_\text{iso}$ correlations. To confirm that, we investigated a possible evolution of $E_\text{p,i}$ -- $E_\text{iso}$ correlation with redshift for type II bursts (see Section~\ref{sec:corr_evo}).

\section{The $\bmath{E}_\text{p,i}$ -- $\bmath{E}_\text{iso}$ correlation}
\label{sec:corr}

\subsection{Fitting the correlation}
\label{sec:corr_an}

\begin{table*}


 \caption{The $E_\text{p,i}$ -- $E_\text{iso}$ correlation fit parameters for investigated subsamples.}  \label{ta_par_am}

 \begin{tabular}{lccccccccc}
  \hline

Subsample	&	N	&	$\rho$	&	$P_{\rho}$	&	$a_\text{York}$			&	$b_\text{York}$	& $a_\text{Deming}$	& $b_\text{Deming}$ 	&	$a$			&	$b$	\\ \hline

I	& 	45	&	0.80	&	4.6 $\times~ 10^{-11}$	&	0.45	$\pm$	0.03	&	0.86	$\pm$	0.04	& 0.37 $\pm$ 0.04	& 0.81	$\pm$	0.05	&	0.41	$\pm$	0.05	&	0.83 $\pm$ 0.06	\\	
I + EE	&	11	&	0.83	&	1.7 $\times~ 10^{-3}$	&	0.40	$\pm$	0.07	&	0.80	$\pm$	0.05	& 0.36 $\pm$ 0.10 &	0.83	$\pm$	0.10	&	0.38	$\pm$	0.12	&	0.82 $\pm$ 0.11	\\	
I w/o EE	&	34	&	0.81	&	7.1 $\times~ 10^{-9}$	&	0.45	$\pm$	0.03	&	0.88	$\pm$	0.05	& 0.37 $\pm$ 0.04 &	0.80 $\pm$ 0.06	&	0.41	$\pm$	0.05	&	0.84 $\pm$ 0.08	\\	\hline
II	&	275	&	0.77	&	< 1 $\times~ 10^{-50}$	&	0.47	$\pm$	0.02	&	-0.36	$\pm$	0.04	& 0.38 $\pm$ 0.02  &	-0.12	$\pm$	0.04	&	0.43	$\pm$	0.03	&	-0.24 $\pm$ 0.06	\\	
II + SN	&	40	&	0.73	&	8.8 $\times~ 10^{-8}$	&	0.57	$\pm$	0.04	&	-0.58	$\pm$	0.10	& 0.34 $\pm$ 0.05 &	-0.13	$\pm$	0.08	&	0.46	$\pm$	0.07	&	-0.36 $\pm$ 0.13	\\	
II w/o SN	&	235	&	0.73	&	9.0 $\times~ 10^{-40}$	&	0.41	$\pm$	0.02	&	-0.23	$\pm$	0.05	& 0.39 $\pm$ 0.03 &	-0.12	$\pm$	0.06	&	0.40	$\pm$	0.03	&	-0.18 $\pm$ 0.07	\\	\hline
II $z_1$	&	55	&	0.69	&	6.2 $\times~ 10^{-9}$	&	0.58	$\pm$	0.04	&	-0.56	$\pm$	0.09	& 0.33 $\pm$ 0.05&	-0.06 $\pm$ 0.07	&	0.45	$\pm$	0.06	&	-0.31 $\pm$ 0.11	\\	
II $z_2$	&	55	&	0.76	&	1.4 $\times~ 10^{-11}$	&	0.42	$\pm$	0.03	&	-0.29	$\pm$	0.08	& 0.37 $\pm$ 0.05 &	-0.14	$\pm$	0.09	&	0.39	$\pm$	0.05	&	-0.22 $\pm$ 0.12	\\	
II $z_3$	&	55	&	0.67	&	1.7 $\times~ 10^{-8}$	&	0.44	$\pm$	0.04	&	-0.31	$\pm$	0.13	& 0.48 $\pm$ 0.07 &	-0.32	$\pm$	0.16	&	0.46	$\pm$	0.09	&	-0.32 $\pm$ 0.21	\\	
II $z_4$	&	55	&	0.73	&	2.1 $\times~ 10^{-10}$	&	0.34	$\pm$	0.05	&	-0.03	$\pm$	0.13	& 0.41 $\pm$ 0.05 &	-0.19	$\pm$	0.13	&	0.38	$\pm$	0.07	&	-0.11 $\pm$ 0.18	\\	
II $z_5$	&	55	&	0.69	&	4.3 $\times~ 10^{-9}$	&	0.40	$\pm$	0.04	&	-0.12	$\pm$	0.11	& 0.46 $\pm$ 0.06 &	-0.28	$\pm$	0.14	&	0.43	$\pm$	0.07	& -0.20 $\pm$ 0.18	\\	\hline

 \end{tabular}
\end{table*}

\begin{figure}
	\includegraphics[width=1.0\columnwidth]{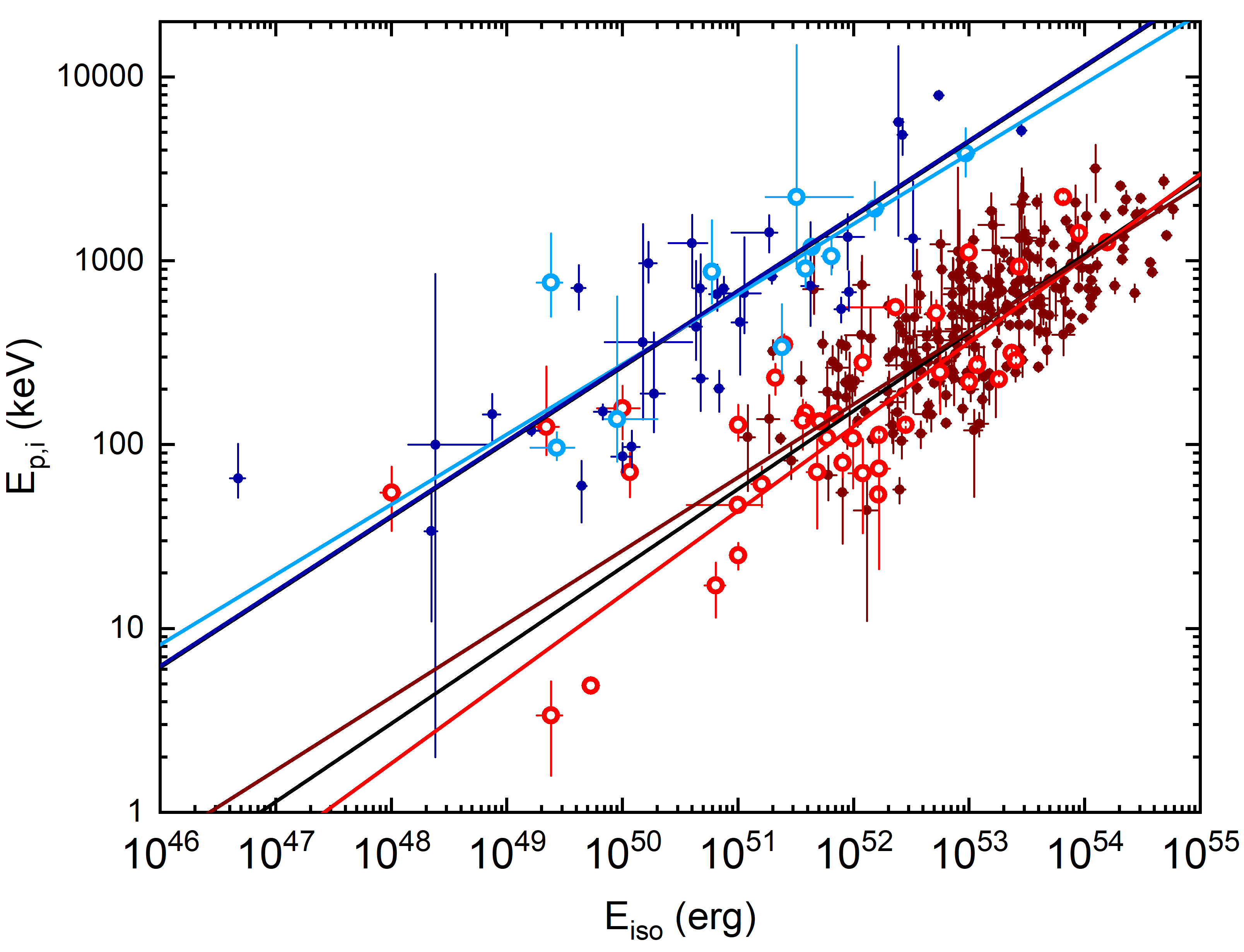}
    \caption{The $E_\text{p,i}$ -- $E_\text{iso}$ correlation for subsamples of regular type I bursts (blue dots), type I bursts with an extended emission (cyan circles), regular type II bursts (dark red dots) and type II bursts with an associated supernova (red circles). Corresponding fits are shown by the colored lines (black lines represent fits of the whole type I and type II samples).
    Uncertainties for $E_\text{iso}$ and $E_\text{p,i}$ are presented at the 1 $\sigma$ significance level.}
    \label{fig:jointfit}
\end{figure}

The $E_\text{p,i}$ -- $E_\text{iso}$ diagram constructed for different subsamples (regular type I bursts, I+EE bursts, regular type II bursts, II+SN bursts) is presented at Fig.~\ref{fig:jointfit}.

To confirm the existence of correlations we calculate the Spearman rank-order correlation coefficients ($\rho$) and the associated null-hypothesis (chance) probabilities or $P_{\rho}$ values (Table~\ref{ta_par_am}). For whole type I and type II samples correlation parameters are $\rho$ = 0.80 ($P_{\rho}$ = 4.6 $\times~ 10^{-11}$) and $\rho$ = 0.77 ($P_{\rho}$ < 1 $\times~ 10^{-50}$), correspondingly, indicating a strong correlation for both samples. The correlation for type I bursts was confirmed for the first time with such high probability level. A strong correlation is also found for II+SN subsample ($\rho$ = 0.73 and $P_{\rho}$ = 8.8 $\times~ 10^{-8}$), while the correlation of I+EE subsample is not highly significant ($\rho$ = 0.83 and $P_{\rho}$ = 1.7 $\times~ 10^{-3}$), due to the small number of events.

The correlation is characterized by the ``intrinsic'' scatter for all subsamples: the observed scatter of the correlation is sufficiently larger than the other one connected with statistical errors only \citep[see e.g.][]{heus13,tsv17}. The intrinsic scatter dominates on the statistical one for the brightest bursts (see e.g. type II bursts with $E_\text{iso} > 10^{54}$ erg at Fig.~\ref{fig:jointfit}). Its presence is confirmed by the $\chi^2_\text{red}$ $\gg$ 1. The nature of the intrinsic scatter is unknown, it could be connected with underestimated systematics (e.g. obtained during spectral fitting and determining of $E_\text{p}$) or with GRB progenitor features. It makes GRBs not standard candles, even when using the $E_\text{p,i}$ -- $E_\text{iso}$ correlation for ``standardizing'' them.

To fit the correlation (equation~(\ref{eq:amati})), we transform $E_\text{p,i}$ and $E_\text{iso}$ values to $x$ = $\lg\Big(\frac{E_\text{iso}}{10^{51}~\text{erg}}\Big) $ and $y$ = $\lg\Big(\frac{E_\text{p,i}}{100~\text{keV}}\Big)$ and perform linear model fitting ($y$ = $ax + b$) with considering of statistical errors for both $x$ and $y$ values using several different methods. Fitting the correlation is appeared to be tricky because of the intrinsic scatter presence.

\begin{equation}
    \lg\Big(\frac{E_\text{p,i}}{100~\text{keV}}\Big) = a\lg\Big(\frac{E_\text{iso}}{10^{51}~\text{erg}}\Big) + b.
	\label{eq:amati}
\end{equation}

Firstly, we use the $\chi^2 = \sum_{i=1}^N\frac{(y_i - ax_i - b)^2}{a^2\sigma_{x_i}^2 + \sigma_{y_i}^2 + \sigma_\text{int}^2}$ minimizing method, which introduces additional intrinsic scatter component $\sigma_\text{int}$ in the weights calculation \citep{trem02}). The method was used for the $E_\text{p,i}$ -- $E_\text{iso}$ correlation fitting in \citet{tsv17}. The intrinsic scatter $\sigma_\text{int}$ is estimated from the $\chi^2_\text{red}$ = 1 condition.

Introducing $\sigma_\text{int}$ in the fitting changes results significantly comparing with $\sigma_\text{int}$ = 0 situation. We found the fit results with $\sigma_\text{int}$ to be similar with the fitting without considering errors at all. Indeed, introducing $\sigma_\text{int}$ equalizes the fit weights, it leads to significant weight decrease for bursts with high S/N. Despite the fact, that both $\sigma_{x}$ and $\sigma_{y}$ errors are included in the approximation, the method measures the deviation of the data against the model in direction parallel to the $y$ axis. As a consequence, fit parameters change dramatically with $x$ $\rightarrow$ $y$ and $y$ $\rightarrow$ $x$ replacing (e.g. for the whole type II sample the power-law index changes from $a$ = 0.35 $\pm$ 0.02 to $a$ = 0.55 $\pm$ 0.03). Therefore, we do not use this method to fit the correlation.

We found York \citep{york04} and Deming \citep{deming11} approximation methods to be reliable, considering both $\sigma_{x}$ and $\sigma_{y}$ errors, not sensitive to $x$ $\rightarrow$ $y$ and $y$ $\rightarrow$ $x$ replacing. We used them both, because they give slightly different parameter values. As a final fit, used for further analysis, we suggest mean values of the fit parameters, $a$ = $\frac{1}{2}$($a_\text{York}$+$a_\text{Deming}$) and $b$ = $\frac{1}{2}$($b_\text{York}$+$b_\text{Deming}$). Corresponding parameter errors are calculated as $\sigma_{a}$ = $\sqrt{\sigma_{a_\text{York}}^2+\sigma_{a_\text{Deming}}^2}$ and $\sigma_{b}$ = $\sqrt{\sigma_{b_\text{York}}^2+\sigma_{b_\text{Deming}}^2}$.

\subsection{The approximation results}
\label{sec:corr_joint}

The approximation results are summarized in Table~\ref{ta_par_am} and shown at Fig.~\ref{fig:jointfit}. Power-law indexes ($a$) of the correlation are found to be the same (within 1 $\sigma$) for all investigated subsamples, $a$ $\simeq$ 0.4, while $b$ values differ significantly: the correlation for type I bursts is shifted up against the one for type II bursts. These features could be used in a blind GRB classification (see Section~\ref{sec:corr_class}). The fact of similar power-law index value for both GRB types confirms results obtained in \citet{zha18a}.

Type I bursts with an extended emission and regular type I bursts follow the same correlation, possibly indicating the same progenitors and emission mechanism. We obtain similar results for type II bursts associated with Ic supernovae and for regular type II bursts.

One of possible explanations of the $E_\text{p,i}$ -- $E_\text{iso}$ correlation is connected with viewing angle effects. In the model, the value of the power-law index depends on the structure of the ejecta: e.g. $E_\text{p,i} \sim E_\text{iso}^{~1/3}$ if it is a cone relativistic jet emission, and  $E_\text{p,i} \sim E_\text{iso}^{~1/4}$ if it is a spherical relativistic emission \citep[e.g.][]{eic04,lev05,poz18a}. The derived value of the power-law index, $a$ $\simeq$ 0.4 is close to the cone relativistic jet emission. The same power-law index for both GRB types can indicate the same structure of the jet for both GRB types.

In the model of connection of $E_\text{p,i}$ -- $E_\text{iso}$ correlation with viewing angle effects, bursts observed close to axis are brighter and harder, placed at top-right in the $E_\text{p,i}$ -- $E_\text{iso}$ plane, while the bursts seen off-axis are placed in bottom-left of the plane. GRB 170817A, associated with gravitational wave event GW 170817, is the principal candidate for an off-axis burst, being the faintest GRB with $E_\text{iso}$ < $10^{47}$ erg (Fig.~\ref{fig:jointfit}). Several signs of observing this burst at a large angle to the jet axis were actually discovered, supporting the viewing angle model. One can mention the absence of the X-ray afterglow at early stages and a powerful kilonova component in optical range, comparing with a standard optical a power-law like afterglow emission \citep[e.g.][]{poz18a}.

\subsection{Outliers of the correlation for type II bursts}
\label{sec:corr_out}

Three type II bursts are found to be placed outside a 3 $\sigma_\text{cor}$ correlation region: GRB 980425B, GRB 031203A, and GRB 171205A. We consider them as outliers of the correlation. The $\sigma_\text{cor}$ corresponds to the ``real'' scatter of the correlation, representing both statistical and intrinsic scatter and estimated as empirical standard deviation of the points from the correlation: $\sigma_\text{cor} = \sqrt{\frac{\sum_{i=1}^n(y_i-\bar{y_i})^2}{n-1}}$, where $\bar{y_i} = ax_i + b $. We also fit the $E_\text{p,i}$ -- $E_\text{iso}$ correlation for type II bursts sample and its subsamples with outliers excluded and find the parameters of the correlation to be similar (within 1 $\sigma$) with ones obtained for subsamples with outliers included.

All three bursts are known outliers and are placed above the $E_\text{p,i}$ -- $E_\text{iso}$ correlation. Their peculiarity is still a subject of debate \citep[e.g.][]{ghi06,dai07,nav12,delia18}, with the following possible interpretations: 1) a normal (intrinsically bright) gamma-ray burst seen off-axis, 2) the burst with different emission mechanism, or 3) an intrinsically weak gamma-ray burst seen on-axis through a scattering screen. These bursts are among the faintest over the investigated type II sample, having $E_\text{iso} \leq 10^{50}$ erg. Excluding them, there are only two type II bursts with $E_\text{iso} \leq 10^{50}$ erg, both are X-ray flashes (XRFs), GRB 020903 and GRB 060218. It leads us to several questions, which should be answered in future works. 1) How is the sample of the faintest ($E_\text{iso} < 10^{50}$ erg) and closest as a consequence ($z \leq 0.1$) bursts biased? 2) Do the closest bursts ($z \leq 0.1$) belong to the different progenitor population? 3) Do the faintest bursts have a different emission mechanism?

To conclude, the nature of the outliers and the behavior of the correlation for the faintest type II bursts ($E_\text{iso} < 10^{50}$ erg) remain unresolved. Increasing the sample of the faintest bursts could shed light on a problem.

\subsection{The evolution of the $\bmath{E}_\text{p,i}$ -- $\bmath{E}_\text{iso}$ correlation with redshift for type II bursts}
\label{sec:corr_evo}

The extensive sample of type II bursts (275 events) allows us to investigate the evolution of the $E_\text{p,i}$ -- $E_\text{iso}$ correlation. We formed five subsamples ($z_1$ -- $z_5$).

\begin{figure}
	\includegraphics[width=1.0\columnwidth]{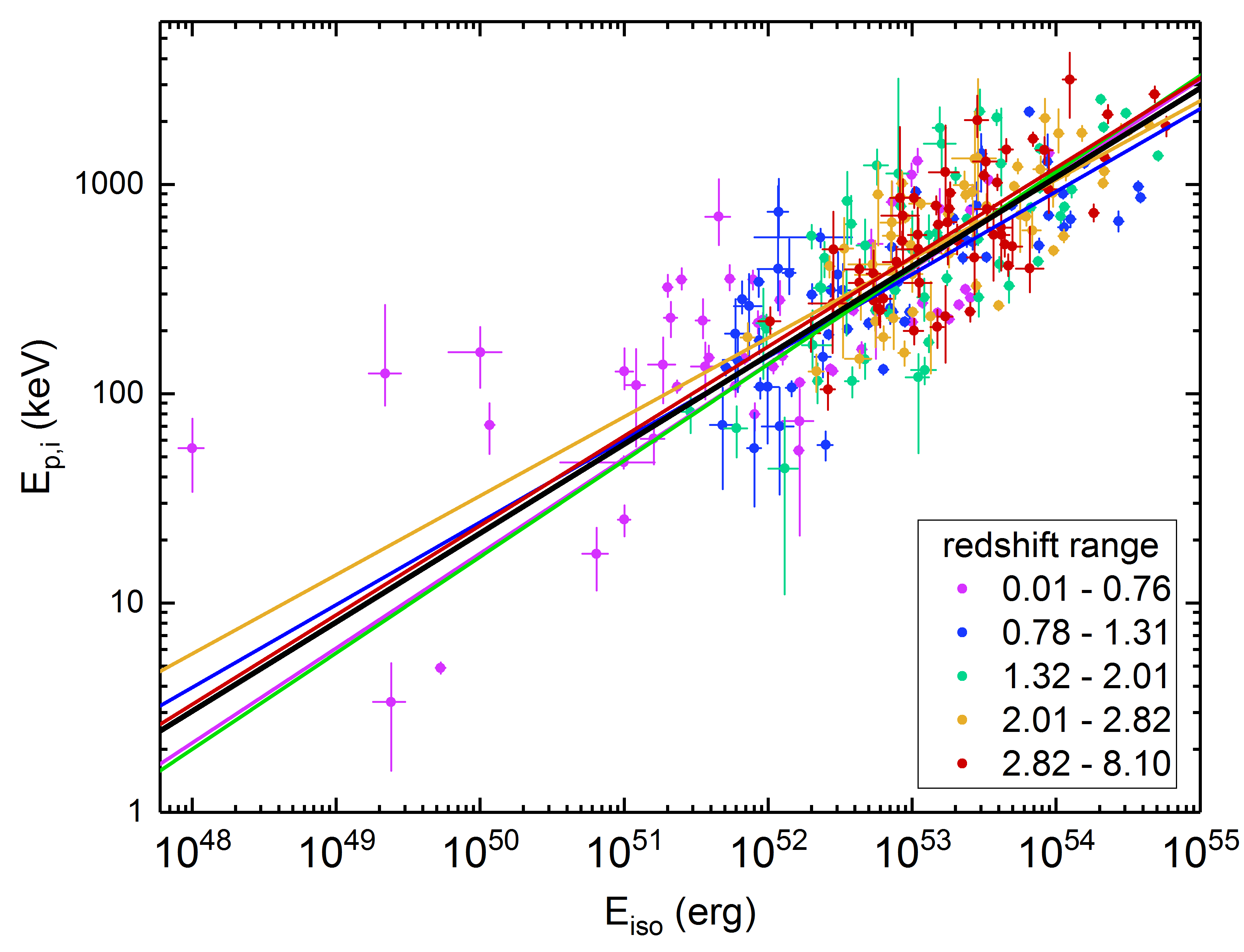}
    \caption{The $E_\text{p,i}$ -- $E_\text{iso}$ correlation for five subsamples of type II bursts (magenta is for $z_1$ subsample, blue -- $z_2$, green -- $z_3$, orange -- $z_4$, red -- $z_5$). Redshift intervals for the subsamples are shown in the legend. Uncertainties for $E_\text{iso}$ and $E_\text{p,i}$ are presented at the 1$\sigma$ significance level. Corresponding fits for the whole sample (black line) and for the subsamples (colored lines) are shown. }
    \label{fig:evofit}
\end{figure}

The $E_\text{p,i}$ -- $E_\text{iso}$ correlation for $z_1$ -- $z_5$ subsamples and corresponding power-law fits are presented in Fig.~\ref{fig:evofit} and summarized in Table~\ref{ta_par_am}. We do not detect the significant evolution of the correlation with the redshift. All subsamples have similar (within 1 $\sigma$) power-law index, $a$ $\simeq$ 0.4. It indicates, that the correlation is not connected with selection effects and confirms negligible selection effects in measuring of $E_\text{p,i}$, discussed in Section~\ref{sec:sample_epi}.

\section{New classification scheme}
\label{sec:corr_class}

The motivation of introducing new classification scheme is connected with phenomenological differences between type I and type II bursts, discussed in the paper.

\subsection{Classification parameters $\bmath{EH}$ and $\bmath{EHD}$}
\label{sec:corr_class_par}

The $E_\text{p,i}$ -- $E_\text{iso}$ correlation for type I bursts sample is found to be well-distinguished from the one constructed for type II bursts and has a similar power-law index value, $E_\text{p,i}$ $\sim$ $E_\text{iso}^{~0.4}$ (Fig.~\ref{fig:jointfit}). Therefore, it can be used for a blind classification of GRBs: the distribution of burst deviations from the power-law dependence $E_\text{p,i}$ $\sim$ $E_\text{iso}^{~0.4}$ should be bimodal. The deviation could be measured by the parameter $EH$ (``Energy-Hardness'') -- the combination of $E_\text{p,i}$ and $E_\text{iso}$ parameters (Equation~\ref{eq:EH}). Type I bursts are harder (higher value of Ep,i) and fainter (lower value of Eiso)  than type II ones in general, therefore they are characterized by a higher value of EH parameter. The $E_\text{p,i}$ value alone is analogous to the commonly used hardness ratio (HR) parameter.

\begin{equation}
    EH = \frac{(E_\text{p,i} / 100~\text{keV})}{ (E_\text{iso} / 10^{51}~\text{erg})^{~0.4}}.
	\label{eq:EH}
\end{equation}

The second proposed classification parameter follows the bimodal nature of duration distribution: type I bursts are shorter than type II ones. Although $T_\text{90,i}$ value is highly affected by selection effects and some intrinsic properties, it can be a good additional classification indicator.

\begin{figure}
	\includegraphics[width=\columnwidth]{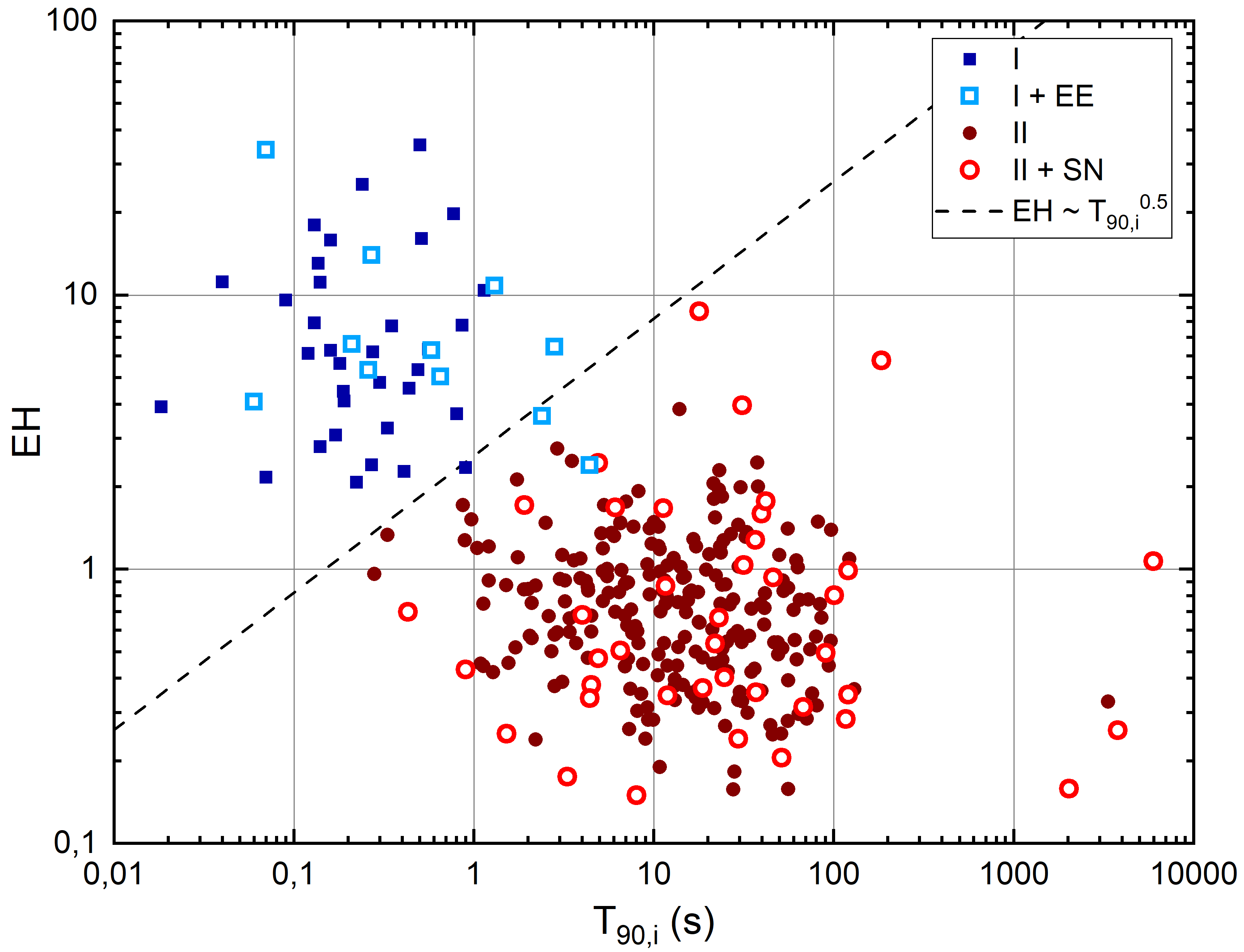}
    \caption{The $T_\text{90,i}$ -- $EH$ diagram for I+EE bursts (cyan open squares), regular type I bursts (blue squares), II+SN bursts (red open circles) and type II (dark red circles) bursts. The power-law dependence with index of 0.5 (dashed line) represents the clusters separation.}
    \label{fig:EHT_diagram}
\end{figure}

The $T_\text{90,i}$ -- $EH$ diagram is presented at Figure~\ref{fig:EHT_diagram}. Type I bursts are placed at the top left on the diagram being short hard and faint, comparing with type II bursts. The separation of the clusters is not satisfactory, when using only $T_\text{90,i}$ (separation, parallel to y-axis) or $EH$ (separation, parallel to x-axis) parameters. Good separation is reached with the dashed line, $EH \sim T_\text{90,i}^{~0.5}$. Therefore, we can modify the $EH$ parameter by including the duration $T_\text{90,i}$ into a consideration and obtain the second classification parameter, $EHD$ (``Energy-Hardness-Duration'', equation~\ref{eq:EHD}). Purely phenomenological, $EHD$ parameter results in the best separation of type I bursts from type II ones.

\begin{equation}
    EHD = \frac{EH}{(T_\text{90,i} / 1~\text{s})^{~0.5}} = \frac{(E_\text{p,i} / 100~\text{keV})}{ (E_\text{iso} / 10^{51}~\text{erg})^{~0.4} ~ (T_\text{90,i} / 1~\text{s})^{~0.5}}.
	\label{eq:EHD}
\end{equation}

Dividing the $EH$ parameter by $T_\text{90,i}^{~0.5}$ for type II bursts, having $T_\text{90,i}$ $\gg$ 1 s, leads to lower values of $EHD$ for them and better separation from type I bursts.

The investigation of a physical meaning of EH and EHD parameters is beyond the scope of the paper, but they might be due to principal differences of progenitors: e.g., central engine of type I bursts could generate harder emission with higher Lorentz factor, but operate for shorter time, comparing with type II bursts.

\subsection{Testing the reliability of different classification schemes}
\label{sec:corr_class_distr}

To investigate the reliability of the GRB classification, based on observational parameters $EH$, $EHD$ and $T_\text{90,i}$, we analyze corresponding parameter distributions, presented at Fig.~\ref{fig:class}.

\subsubsection{Fitting the distributions}

We perform a blind approximation of distributions. It suggests the approximation of the total (type I + type II samples) parameter distribution by sum of two functions, representing two types of bursts.

\begin{figure*}
	\includegraphics[width=2.0\columnwidth]{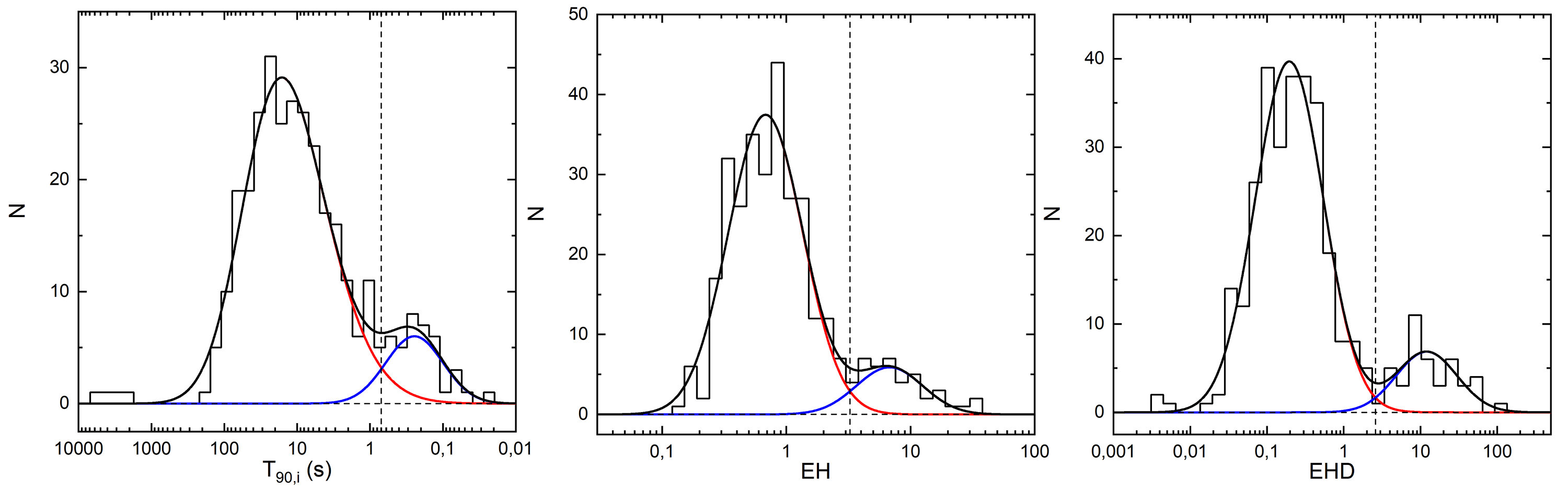}
    \caption{The distributions of the $T_\text{90,i}$ (left),  $EH$ (middle), and $EHD$ (right) parameters for whole samples of type I and type II bursts. Blue (red) curves represent distribution fits for type I (type II) burst modes, while black curves show total fit. The intersection points of type I and type II modes fits are shown by vertical dashed lines.}
    \label{fig:class}
\end{figure*}

The duration ($T_\text{90,i}$) distribution is fitted in a logarithmic scale by sum of gaussian and skewed gaussian (equation~\ref{eq:gauss}), representing type I and type II bursts, correspondingly. Using skewed gaussian instead of regular gauss function for type II burst mode is connected with deformation of the distribution at the left side (large $T_\text{90,i}$ values) probably due to several selection effects, discussed earlier (see Section~\ref{sec:sample_prop_dur}). We fix the skew parameter in the fitting, $\alpha$ = -1.5. The equation~\ref{eq:gauss} with $\alpha$ = 0 corresponds to the regular gaussian. The possibility of using skewed gaussian to fit duration distribution is also discussed e.g. in \citet{tarno16}. Other distributions ($EH$ and $EHD$) are fitted by sum of two regular gaussians ($\alpha$ = 0) in a logarithmic scale. Fit results are summarized in Table~\ref{ta_distr} and shown in Fig.~\ref{fig:class}.

\begin{equation}
   N = \frac{A}{\sqrt{2\pi\omega^2}}\exp\Big(-\frac{(x - x_0)^2}{2\omega^2}\Big)\Big[1 + {\rm erf}\Big(\frac{\alpha(x - x_0)}{\sqrt{2\omega^2}}\Big)\Big].
	\label{eq:gauss}
\end{equation}

\begin{table}


 \caption{Parameters of investigated distributions.}  \label{ta_distr}

 \begin{tabular}{lccccc}
  \hline

Distr.	        &	GRB	&	A			    &	x$_0$			        &	$\omega$			&	$\alpha$	\\\hline
$T_\text{90,i}$	&	 I	&	5.9	$\pm$	0.7	&	-0.61	$\pm$	0.05	&	0.39	$\pm$	0.05	&	0	\\	
	            &  	 II	&	42.1 $\pm$	0.8	&	1.64	$\pm$	0.01	&	0.79	$\pm$	0.02	&	-1.5	\\	
$EH$	       &	 I	&	4.0	$\pm$	0.5	&	0.83	$\pm$	0.03	&	0.27	$\pm$	0.04	&	0	\\	
	           &	 II	&	28.3 $\pm$	0.5	&	-0.16	$\pm$	0.01	&	0.30	$\pm$	0.01	&	0	\\	
$EHD$		   &	 I	&	6.8	$\pm$ 0.6	&	1.08	$\pm$	0.04	&	0.40	$\pm$	0.04	&	0	\\	
	       	   &	 II	&	44.3 $\pm$	0.6	&	-0.71	$\pm$	0.01	&	0.45	$\pm$	0.01	&	0	\\	
	\hline

 \end{tabular}
\end{table}

\subsubsection{Testing the different classification methods reliability}

Traditional GRB blind classification method, based on the duration distribution in the observer frame, uses the position of the distribution minimum, observed at $T_\text{90}$ $\sim$ 2 s \citep[e.g.][]{kou93}. In the rest frame the position of the minimum is shifted towards short durations ($T_\text{90,i}$ distribution). It is close to the intersection point (dashed vertical line at Fig.~\ref{fig:class}, left) of individual modes of two burst types (blue and red dashed curves at Fig.~\ref{fig:class}, left). The intersection point is placed at $T_\text{90,i}$ = 0.7 s in our sample. We use the intersection point as the separation line in a blind classification for all investigated distributions.

As seen at Fig.~\ref{fig:class}, modes representing type I and type II samples in the duration distribution, are highly overlapped. The intersection point is placed at a half of the height of type I distribution, the contribution of type II mode is significant down to $T_\text{90,i}$ $\sim$ 0.2 s. We can estimate the distributions overlap numerically by calculating the portion of the type I (type II) bursts with $T_\text{90,i}$ larger (smaller) than the intersection value (0.7 s), using the corresponding model functions derived in the blind fitting procedure (Table~\ref{ta_distr}). While the most of type II bursts (97.7 \%) are within the interval $T_\text{90,i}$ > 0.7 s, more than 12 \% of type I bursts are outside the $T_\text{90,i}$ < 0.7 s region. These values correspond to the expected portion of falsely classified bursts. It makes the reliability of $T_\text{90,i}$ based classification low.

To check the reliability of the $T_\text{90,i}$ based classification method, we perform a blind classification of bursts from our sample. 9 type I bursts (20 \% of the sample) and 3 type II bursts (1 \% of the sample) are classified falsely, having $T_\text{90,i}$ larger or smaller than 0.7 s, correspondingly (Table~\ref{ta_class}). The portion of falsely classified bursts roughly follows the expected value, based on the overlap of the distribution. I+EE bursts are the most significant outliers in the duration distribution with GRB 060614A being the ``leader'' ($T_\text{90,i}$ = 4.4 s).

We perform the same analysis for $EH$ and $EHD$ distributions. The results are summarized in Table~\ref{ta_class}. The $EH$ distribution gives better results, comparing with the $T_\text{90,i}$ one. The separation point is placed at $EH$ = 3.3, which gives comparable overlapping: 12 \% of type I and 1.1 \% of type II bursts are placed beyond the separation point. 9 type I bursts and 4 type II bursts are falsely classified using $EH$-based blind classification scheme. The most significant outliers in type II sample are known outliers of the $E_\text{p,i}$ -- $E_\text{iso}$ correlation (GRB 980425B, GRB 031203 and GRB 171205A, see Section~\ref{sec:corr_out}). Among 9 type I bursts, falsely classified, only GRB 060614A represents the subsample of I+EE bursts.

The best results are obtained for $EHD$ distribution. The separation point is placed at $EHD$ = 2.6 with almost negligible overlapping: 6.5 \% of type I and 0.6 \% of type II bursts are placed beyond the separation point. Only three type I bursts (GRB 050724A, GRB 060614A and GRB 131004A) are falsely classified using $EHD$-based blind classification scheme. All type II bursts are classified correctly.

GRB 050724A is classified correctly in $EH$-based scheme, being outlier in duration-based scheme. It is not significant outlier in $EHD$-based scheme, having $EHD$ = 2.33 with $EHD$ = 2.6 used for classification. It may indicate, that its false classification could be accidental. Indeed, the $E_\text{p,i}$ parameter is determined for the burst with large errors: $E_\text{p,i}$ = 138 $_{-57}^{+503}$ keV (Table~\ref{ta1}). Changing $E_\text{p,i}$ to 200 keV results in $EHD$ = 3.4, making the burst to be from type I population.

Falsely classified in all three classification schemes GRB 060614A is a known outlier, demonstrating observational features of type II bursts, except the absence of supernova component \citep[e.g.][]{geh06}. Nevertheless, it probably belongs to I+EE population, confirmed by a marginal detection of a kilonova emission, characteristics of type I bursts \citep{yang15}. The possible solution of its peculiarity is discussed in the Section~\ref{sec:corr_class_traj}.

GRB 131004A is also classified falsely in all three classification schemes, but the discrepancy is not highly significant. The $E_\text{p,i}$ value was not measured precisely ($E_\text{p,i}$ = 202 $\pm$ 51 keV), so the false classification could be accidental. Therefore, we can not exclude the burst as the type I one.

As was mentioned before, GRB 170817A (the first gamma-ray burst associated with gravitational waves) is placed in the middle of the overlap region of hardness ratio and duration distributions \citep[e.g.][]{gol17}, confusing its blind classification. But the burst is clearly classified as the type I burst in our both, $EH$ and $EHD$ classification methods, emphasizing their high reliability.

To conclude, $EHD$-based classification method is the most reliable one for a blind classification of GRBs. The values of $EH$ and $EHD$ parameters along with the GRB type based on the corresponding blind classification are presented in Table~\ref{ta1} for all bursts.

\begin{table}


 \caption{The reliability of different blind classification schemes.}  \label{ta_class}

 \begin{tabular}{lccc}
  \hline

Distribution		&	Type I $^a$	&	Type II	$^a$  & Separation $^b$\\	\hline
$T_\text{90,i}$	&	12.4\% (9) 	&	2.3\% (3)	& 0.7 s \\	
$EH$	     	&	12.1\% (9)	    &	1.1\% (4)	& 3.3 \\	
$EHD$	    	&   6.5\% (3)	   &	0.6\% (0)	& 2.6 \\	
\hline

\multicolumn{4}{l}{$^a$ the expected portion of bursts beyond the separation line with the}\\
\multicolumn{4}{l}{real number of false blind classifications for the sample in brackets.}\\
\multicolumn{4}{l}{$^b$ the position of the separation line.}\\

 \end{tabular}
\end{table}

\subsection{Using $\bmath{EHD}$-based classification scheme for GRBs with no redshift}
\label{sec:corr_class_traj}

Although $EHD$-based classification gives the most reliable results, it is based on redshift and $E_\text{p}$ values, which are obtained for a very small portion (< 10 \%) of observed gamma-ray bursts. Let us investigate the dependence of $EH$ and $EHD$ parameters on redshift.

\begin{figure}
	\includegraphics[width=\columnwidth]{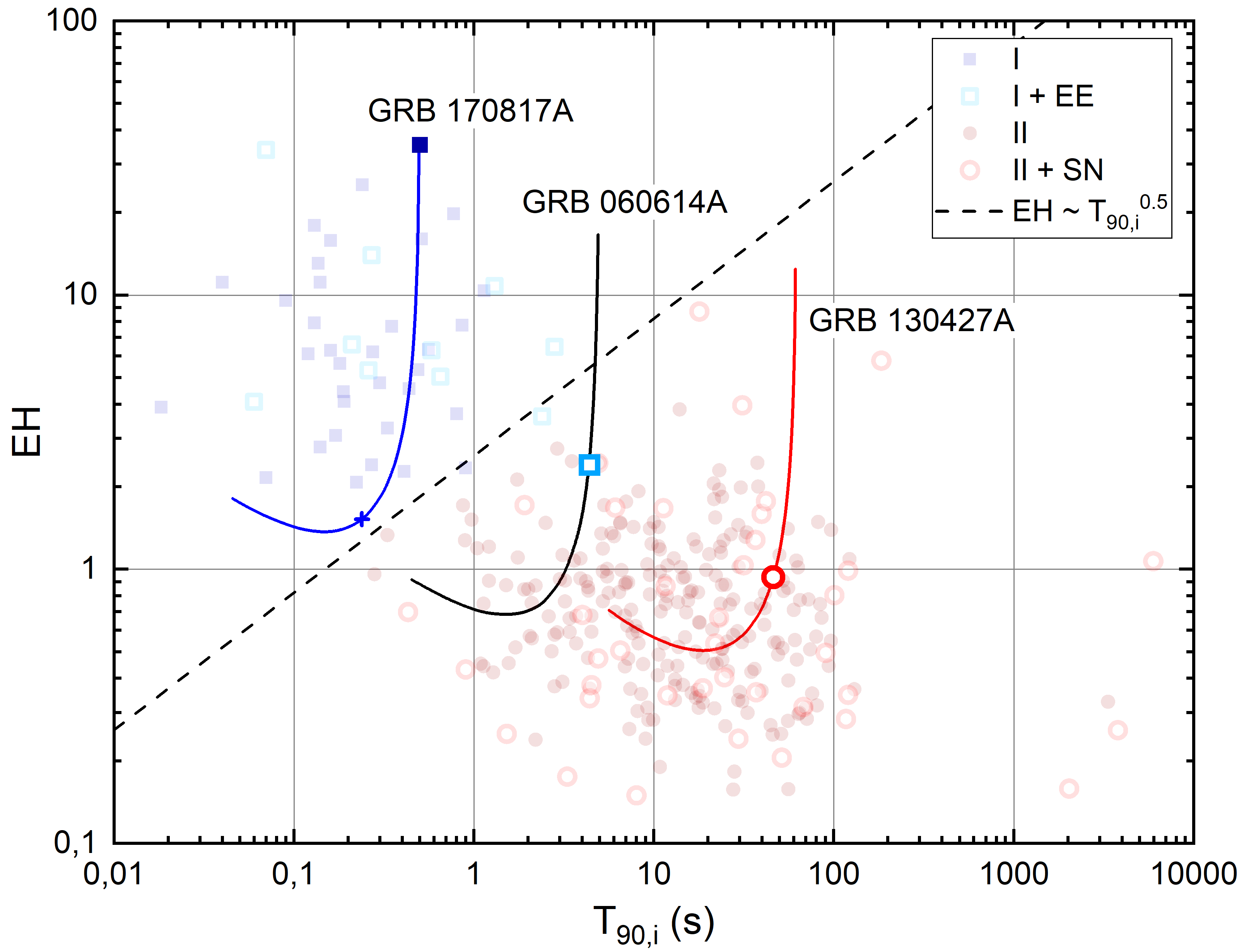}
    \caption{The $T_\text{90,i}$ -- $EH$ diagram. The trajectories, representing the dependence of parameters on redshift, are shown for GRB 060614A by black curve, for GRB 170817A by blue curve, for GRB 130427A by red curve. The closest point of the blue curve to the separation line $EHD$ = 2.6 (dashed line) is shown by plus symbol. The diagram for I+EE bursts (cyan open squares), regular type I bursts (blue squares), II+SN bursts (red open circles) and type II (dark red circles) bursts is shown by semi-transparent symbols, except GRB 060614A, GRB 170817A and GRB 130427A.}
    \label{fig:EHT_diagram_tr}
\end{figure}

The trajectory of GRB 060614A on $T_\text{90,i}$ -- $EH$ diagram, as a function of redshift in the interval $z$ = (0.01, 10), is shown by a black curve in Fig.~\ref{fig:EHT_diagram_tr}. The measured redshift for the burst ($z$ = 0.1254) corresponds to the cyan square symbol and characterizes the burst as the type II one. The trajectory crosses the classification line ($EHD$ = 2.6, dashed line at Fig.~\ref{fig:EHT_diagram_tr}) at $z$ $\simeq$ 0.03. So, the burst could be classified as the type I one, if $z$ < 0.03.

It is possible, that redshift of GRB 060614A is determined wrong. It is measured only for a supposed host galaxy, placed 0.5$\arcsec$ from the burst optical counterpart. The estimated chance probability of the observed offset between the GRB and the galaxy ranges from $P$ = 2 $\times$ 10$^{~-5}$ in \citet{geh06} to $P$ $\simeq$ 0.01 in \citet{cob06}. So, the association of the burst with the galaxy could be accidental, and the burst could be placed in outskirts of other nearby galaxy with $z$ < 0.03. This suggestion is also supported by unusually high mass of ejecta of detected kilonova for the event, estimated for $z$ = 0.125 \citep{jin16,tan16}. Placing the burst at lower redshift eliminates the peculiarity.

We investigated the change of the trajectory on $T_\text{90,i}$ -- $EH$ diagram, varying the shape of the GRB gamma-ray spectrum, and found only marginal change, following from the $k$-correction. Two another trajectories, representing famous type I GRB 170817A and type II GRB 130427A bursts, demonstrate the similar behavior (Fig.~\ref{fig:EHT_diagram_tr}).

The blue curve, constructed for GRB 170817A, lies above the separation line $EHD$ = 2.6 in the whole considered range of redshift. Therefore, the burst can be classified as the type I one, even if the redshift is unknown. The closest point of the curve to the separation line is placed at $z$ $\simeq$ 1.1, shown by plus symbol at Fig.~\ref{fig:EHT_diagram_tr}. To conclude, if the value of $EHD$ parameter is greater than 2.6 (the separation value) at $z$ = 1.1 for the investigated burst, one can classify it as the type I burst.

The red curve, constructed for GRB 130427A, is placed below the separation line at any considered redshift, indicating the burst to be from the type II population. So, if the value of $EHD$ parameter is less than the separation value at the bounds of the considered redshift ($z$ = 0.01 and $z$ = 10 in our case), we can classify the burst (with no redshift measured) as the type II one.

In the intermediate region (e.g., as for GRB 060614A), when the trajectory crosses the separation line, we can not perform the classification, but we can restrict the possible redshift range of the source, if we know exactly the type of the burst from some other observational features (e.g. the absence of supernova and the presence of kilonova, as for GRB 060614A).

\subsection{Comparison with several other classification methods based on combinations of $\bmath{E}_\text{p}$, $\bmath{E}_\text{iso}$ and $\bmath{T}_\text{90,i}$}
\label{sec:corr_class_comp}

Attempts of using the combination of ${E}_\text{p}$, ${E}_\text{iso}$, ${T}_\text{90,i}$ parameters and Amati relation as classification discriminators were made in several papers \citep[e.g.][]{lu10, qin13, zha18a, shah15, zou18, poz18a}. Below we briefly compare the several classification schemes.

\citet{lu10} introduced a new classification parameter $\epsilon$ = ${E}_\text{iso,52}$/${E}_\text{p,i,2}^{~k}$, which showed a clear bimodal distribution for $k$ = 5/3. The value of power-law index $k$ was not based on Amati relation and any other theoretical assumption, it was chosen just to minimize the dependence of $\epsilon$ on redshift.

The $\epsilon$ parameter is somewhat analogous to our $EH$ parameter, being the power-law-like combination of ${E}_\text{p}$ and ${E}_\text{iso}$ parameters. As pointed out in \citet{lu10}, it is not obvious physically why the parameter $\epsilon$ (corresponding to $k$ = 5/3) gives a cozy classification scheme. In that sense, the $EH$ parameter has an advantage of being a direct consequence of Amati relation, although the nature of the Amati relation itself is still questionable.

The $\epsilon$ parameter gives analogous to the $EH$ parameter (but worse comparing with the $EHD$) reliability of corresponding blind classification scheme, although the ``problematic'' GRB 060614A was classified correctly using $\epsilon$-based scheme. We should note, values of ${E}_\text{p}$ and ${E}_\text{iso}$ parameters for several common bursts, presented in both, our and \citet{lu10} samples differ significantly (e.g., for GRB 051221A and GRB 060801A). The reason of that discrepancy is not clear, the samples from both papers are mostly compilations of other papers.

\citet{qin13} suggested the classification scheme, based directly on the ${E}_\text{p,i}$ -- ${E}_\text{iso}$ correlation. They used the distribution of logarithmic deviation from the correlation of type II bursts $\sigma$ = $\lg{E}_\text{p,i}$ - $\lg94$ - 0.57$\lg{E}_\text{iso}$ to estimate the separation value. The deviation $\sigma$ was firstly introduced in \citet{ama06} to calculate the dispersion of the correlation of type II bursts. The classification method and the parameter $\sigma$ are analogous to our $EH$ scheme ($\sigma$ $\sim$ $\lg$ $EH$), except the power-law index value of Amati relation, which is $a$ = 0.57 in \citet{qin13} and $a$ = 0.4 in our analysis. The discrepancy of power-law indices is connected at least with the difference of the samples: our sample contains three times more type I bursts and almost twice times more type II bursts. As we have shown in Section~\ref{sec:corr_an}, fit parameters of the correlation also depend on the method used, which could be another reason of the discrepancy.

In \citet{shah15} the classification parameter $R$ = ${E}_\text{p}$/${T}_\text{90}$ was proposed as least affected parameter by the detection threshold of gamma-ray detectors. It was also suggested, that the two GRB types are most distinctively separated in the ${E}_\text{p}$ -- ${T}_\text{90}$ plane. We check the classification scheme, based on the $R$ parameter and its intrinsic equivalent ($R_\text{i}$ = ${E}_\text{p,i}$/${T}_\text{90,i}$) for GRBs from our sample and found it to be similar with regular duration-based classification scheme. The result is connected with the fact, that the distributions of ${E}_\text{p}$ and ${E}_\text{p,i}$ for type I and type II GRBs are significantly overlapped with minor differences and give only marginal improvement in $R$-based scheme comparing with the duration-based one. But if we use instead of the ${E}_\text{p,i}$ parameter the combination of ${E}_\text{p,i}$ and ${E}_\text{iso}$ parameters, following the bimodal nature of ${E}_\text{p,i}$ -- ${E}_\text{iso}$ correlation, we obtain the $EHD = R_\text{i}~E_\text{iso}^{~-0.4}~T_\text{90,i}^{~0.5} =  E_\text{p,i}~E_\text{iso}^{~-0.4}~T_\text{90,i}^{~-0.5}$ parameter, giving the best separation of the type I / type II bursts of our sample.

\section{Conclusions}

We construct and investigate the most extensive (at the moment) sample of 45 type I and 275 type II gamma-ray bursts with a known redshift (both spectroscopic and photometric) and $E_\text{p}$. We discuss several selection effects, leading to the evolution of $T_\text{90,i}$, $E_\text{iso}$ and $E_\text{p,i}$ parameters with redshift.

We confirm strong $E_\text{p,i}$ -- $E_\text{iso}$ correlation for both types of bursts. The correlation for type I bursts is found to be well-distinguished from the one for type II bursts and has similar power-law index value, $E_\text{p,i}$ $\sim$ $E_\text{iso}^{~0.4}$. It possibly indicates the same emission mechanism for both types.

Three possible outliers from the $E_\text{p,i}$ -- $E_\text{iso}$ correlation for type II subsample are confirmed: GRB 980425B, GRB 031203A and GRB 171205A. They represent the subsample of the faintest type II bursts with $E_\text{iso} < 10^{50}$ erg. The nature of their peculiarity and the behavior of the correlation at $E_\text{iso} < 10^{50}$ erg remain unresolved. Increasing the sample of the faintest bursts could shed light on a problem.

We show that type I bursts with an extended emission and regular type I bursts follow the same correlation. It possibly indicates the same progenitor type for both subsamples of type I bursts. Nevertheless, I+EE bursts possibly form a separate sub-class of type I bursts, which is indicated by the features of redshift and duration distributions.

The same $E_\text{p,i}$ -- $E_\text{iso}$ correlation behavior is also obtained for type II bursts with associated Ic supernovae and regular type II bursts, while the redshift, $E_\text{p,i}$ and $E_\text{iso}$ parameters distributions show significant differences. They are interpreted as strong selections effects, connected with observational constraints on supernova detection at a high redshift. It possibly indicates the same progenitor type for both subsamples of type II bursts, leading to the suggestion, that all type II bursts are accompanied by Ic supernovae.

The significant evolution of the $E_\text{p,i}$ -- $E_\text{iso}$ correlation with redshift for type II bursts is not found. It confirms the weak selection effects on the correlation.

Using the $E_\text{p,i}$ -- $E_\text{iso}$ correlation, we suggest two new classification methods by introducing two parameters, $EH$ and $EHD$, representing the combination of the $E_\text{p,i}$, $E_\text{iso}$ and $T_\text{90,i}$ parameters. $EHD$ parameter is found to be the most reliable one for the GRB classification. It also can be used to classify bursts with no redshift or to estimate their redshift.

\section*{Acknowledgements}

Authors thank the anonymous referee for helpful suggestions and improvement of the paper. Authors acknowledge support from RSF grant 18-12-00378.




\bibliographystyle{mnras}
\bibliography{minaev_base} 



\appendix

\section{The sample of gamma-ray bursts}

\defcitealias{tsv17}{1}
\defcitealias{can17}{2}
\defcitealias{qin13}{3}
\defcitealias{ama06}{4}
\defcitealias{fro09}{5}
\defcitealias{min14}{6}
\defcitealias{zha09}{7}
\defcitealias{nor10}{8}
\defcitealias{vol14}{9}
\defcitealias{svi16}{10}
\defcitealias{but07}{11}
\defcitealias{kru15}{12}
\defcitealias{zou18}{13}
\defcitealias{zha18a}{14}
\defcitealias{gcn7205}{15}
\defcitealias{min17}{16}
\defcitealias{can11}{17}
\defcitealias{nav12}{18}
\defcitealias{gcn10040}{19}
\defcitealias{gcn11722}{20}
\defcitealias{gcn11736}{21}
\defcitealias{poz18}{22}
\defcitealias{sel18}{23}
\defcitealias{gcn13120}{24}
\defcitealias{gcn13118}{25}
\defcitealias{gcn13469}{26}
\defcitealias{gcn13460}{27}
\defcitealias{gcn13517}{28}
\defcitealias{gcn13512}{29}
\defcitealias{can14}{30}
\defcitealias{gcn13634}{31}
\defcitealias{gcn13628}{32}
\defcitealias{gcn13736}{33}
\defcitealias{bha16}{34}
\defcitealias{gcn13810}{35}
\defcitealias{gcn14219}{36}
\defcitealias{gcn14207}{37}
\defcitealias{gcn14437}{38}
\defcitealias{gcn14702}{39}
\defcitealias{gcn14808}{40}
\defcitealias{gcn14848}{41}
\defcitealias{gcn14896}{42}
\defcitealias{gcn14882}{43}
\defcitealias{vol17}{44}
\defcitealias{gcn15260}{45}
\defcitealias{gcn15330}{46}
\defcitealias{gcn15499}{47}
\defcitealias{gcn15494}{48}
\defcitealias{gcn15800}{49}
\defcitealias{gcn15889}{50}
\defcitealias{gcn15883}{51}
\defcitealias{gcn15936}{52}
\defcitealias{gcn15923}{53}
\defcitealias{gcn16150}{54}
\defcitealias{gcn16284}{55}
\defcitealias{gcn16269}{56}
\defcitealias{gcn16306}{57}
\defcitealias{gcn16301}{58}
\defcitealias{gcn16374}{59}
\defcitealias{gcn16425}{60}
\defcitealias{pan19}{61}
\defcitealias{gcn16442}{62}
\defcitealias{gcn16495}{63}
\defcitealias{gcn16505}{64}
\defcitealias{gcn16797}{65}
\defcitealias{gcn16900}{66}
\defcitealias{gcn16902}{67}
\defcitealias{gcn16983}{68}
\defcitealias{gcn17055}{69}
\defcitealias{gcn17228}{70}
\defcitealias{gcn17234}{71}
\defcitealias{gcn17319}{72}
\defcitealias{gcn17358}{73}
\defcitealias{gcn17523}{74}
\defcitealias{gcn17731}{75}
\defcitealias{gcn17759}{76}
\defcitealias{gcn18080}{77}
\defcitealias{gcn18198}{78}
\defcitealias{gcn18205}{79}
\defcitealias{gcn18527}{80}
\defcitealias{gcn18524}{81}
\defcitealias{gcn18598}{82}
\defcitealias{gcn18580}{83}
\defcitealias{gcn19109}{84}
\defcitealias{gcn19106}{85}
\defcitealias{gcn19276}{86}
\defcitealias{zha18}{87}
\defcitealias{gcn19773}{88}
\defcitealias{gcn20059}{89}
\defcitealias{gcn20082}{90}
\defcitealias{gcn20111}{91}
\defcitealias{gcn20197}{92}
\defcitealias{gcn20245}{93}
\defcitealias{gcn20308}{94}
\defcitealias{gcn20321}{95}
\defcitealias{gcn20342}{96}
\defcitealias{gcn20456}{97}
\defcitealias{gcn20458}{98}
\defcitealias{gcn20604}{99}
\defcitealias{gcn20686}{100}
\defcitealias{gcn21247}{101}
\defcitealias{gcn21240}{102}
\defcitealias{gcn21298}{103}
\defcitealias{gcn21299}{104}
\defcitealias{poz18a}{105}
\defcitealias{gcn21799}{106}
\defcitealias{gcn22003}{107}
\defcitealias{gcn22096}{108}
\defcitealias{delia18}{109}
\defcitealias{postigo17}{110}
\defcitealias{gcn22272}{111}
\defcitealias{gcn22384}{112}
\defcitealias{gcn22386}{113}
\defcitealias{gcn22546}{114}
\defcitealias{gcn22566}{115}
\defcitealias{gcn22567}{116}
\defcitealias{gcn22823}{117}
\defcitealias{gcn22996}{118}
\defcitealias{gcn23061}{119}
\defcitealias{gcn23240}{120}
\defcitealias{gcn23246}{121}
\defcitealias{gcn23363}{122}
\defcitealias{gcn23424}{123}
\defcitealias{gcn23488}{124}
\defcitealias{gcn23495}{125}
\defcitealias{gcn23601}{126}
\defcitealias{belk19}{127}
\defcitealias{gcn23637}{128}
\defcitealias{gcn23708}{129}
\defcitealias{gcn23983}{130}
\defcitealias{poz19}{131}

\begin{table*}


 \caption{The sample of gamma-ray bursts.}  \label{ta1}

\begin{tabular}{lccccccccc}
  \hline
 GRB    &  $T_\text{90,i}$ $^a$ &  $z$ $^b$    & $E_\text{iso}$ & $E_\text{p,i}$ & Type $^c$ & Experiment $^d$ & Refs $^e$ & $EH$ Type $^f$ & $EHD$ Type $^g$ \\
        &    s    &      &  $10^{51}$ erg  & keV    &  &   &  	&  &  \\
 \hline

970228	&	31.5	&	0.695 	&	12.01	$ \pm $ 	0.93				&	280	$_{-	42	}^{+	66	}$	&	II+SNph	& Konus-Wind	&  \citetalias{	tsv17, can17}	& 1.04 II & 0.19 II \\
970508	&	10.9	&	0.835 	&	6.12	$ \pm $ 	1.30				&	145	$ \pm $ 	43				&	II	&  BeppoSAX & \citetalias{	qin13}	& 0.70 II & 0.21 II \\
970828	&	33.8	&	0.9578 	&	262.2	$ \pm $ 	9.1				&	531	$_{-	43	}^{+	47	}$	&	II	&	Konus-Wind & \citetalias{	tsv17}		& 0.57 II & 0.10 II \\
971214	&	3.6	&	3.418 	&	146.3	$ \pm $ 	6.1				&	791	$_{-	71	}^{+	88	}$	&	II	&	Konus-Wind & \citetalias{	tsv17} & 1.08 II & 0.57 II \\
980326	&	4.5	&	1 PH &	4.82	$ \pm $ 	0.90				&	71	$ \pm $ 	36				&	II+SNph	&	BeppoSAX & \citetalias{	can17,qin13}	& 0.38 II & 0.18 II \\
980425B	&	17.9	&	0.0085 	&	0.0010	$ \pm $ 	0.0002				&	55	$ \pm $ 	21				&	II+SNsp	& BATSE/CGRO	& \citetalias{	can17, ama06} & 8.72 I & 2.06 II \\
\hline

 \end{tabular}
\end{table*}

\begin{table*}


 \contcaption{}  \label{ta1}

 \begin{tabular}{lccccccccc}
  \hline
 GRB    &  $T_\text{90,i}$  &  $z$     & $E_\text{iso}$ & $E_\text{p,i}$ & Type  & Experiment & Refs & $EH$ Type & $EHD$ Type \\
        &    s    &      &  $10^{51}$ erg  & keV    &  &   &  	&   &    \\
 \hline

050922C	&	2.0	&	2.198	&	53	$ \pm $ 	17				&	415	$ \pm $ 	111				&	II	& HETE-2	& \citetalias{	qin13}	& 0.85 II & 0.60 II \\
051008	&	55.4	&	2.77 $_{-0.20}^{+0.15}$	PH&	834	$ \pm $ 	70				&	2074	$_{-	354	}^{+	501	}$	&	II	&	Konus-Wind & \citetalias{tsv17, vol14}	& 1.41 II & 0.19 II \\
051022	&	85.4	&	0.8	 &	490	$ \pm $ 	10				&	794	$_{-	29	}^{+	32	}$	&	II	&	Konus-Wind & \citetalias{	tsv17}	& 0.67 II & 0.07 II \\
051109A	&	11.9	&	2.346 	&	71.1	$ \pm $ 	8.5				&	569	$_{-	141	}^{+	268	}$	&	II	&	Konus-Wind & \citetalias{	tsv17}	& 1.03 II & 0.30 II \\
051221A	&	0.14	&	0.5465	&	9.10	$_{-	1.12	}^{+	1.29	}$	&	677	$_{-	141	}^{+	200	}$	&	I	&	Konus-Wind & \citetalias{	svi16}	& 2.80 II & 7.48 I \\
060115	&	30.8	&	3.53 	&	63	$ \pm $ 	9				&	285	$ \pm $ 	34				&	II	& Swift	& \citetalias{	qin13}	& 0.54 II & 0.10 II \\
\hline

 \end{tabular}
\end{table*}

\begin{table*}


 \contcaption{}  \label{ta1}

 \begin{tabular}{lccccccccc}
  \hline
 GRB    &  $T_\text{90,i}$  &  $z$     & $E_\text{iso}$ & $E_\text{p,i}$ & Type  & Experiment & Refs & $EH$ Type & $EHD$ Type \\
        &    s    &      &  $10^{51}$ erg  & keV    &  &   &  	&   &    \\
 \hline

080411	&	21.1	&	1.0301 	&	238.5	$ \pm $ 	7.7				&	540	$_{-	37	}^{+	43	}$	&	II	&	Konus-Wind & \citetalias{	tsv17}	& 0.61 II & 0.13 II \\
080413A	&	5.2	&	2.433	 &	72	$ \pm $ 	12				&	659	$_{-	148	}^{+	381	}$	&	II	&	Konus-Wind & \citetalias{	tsv17}	& 1.19 II & 0.52 II \\
080413B	&	1.28	&	1.1014 	&	24	$ \pm $ 	3				&	150	$ \pm $ 	30				&	II	&	Swift  & \citetalias{	qin13}	& 0.42 II & 0.37 II \\
080514B	&	2.04	&	1.8 $_{-0.3}^{+0.4}$	PH &	286	$ \pm $ 	13				&	549	$ \pm $ 	34				&	II	&	Konus-Wind & \citetalias{	tsv17}	& 0.57 II & 0.40 II \\
080602	&	23.0	&	1.8204 	&	80	$ \pm $ 	14				&	1128	$_{-	361	}^{+	2084	}$	&	II	& Konus-Wind & 	\citetalias{	tsv17}	& 1.96 II & 0.41 II \\
080603B	&	3.4	&	2.6892	 &	100	$ \pm $ 	17				&	373	$_{-	66	}^{+	137	}$	&	II	&	Konus-Wind & \citetalias{	tsv17}	& 0.59 II & 0.32 II \\
\hline

 \end{tabular}
\end{table*}

\begin{table*}


 \contcaption{}  \label{ta1}

 \begin{tabular}{lccccccccc}
  \hline
 GRB    &  $T_\text{90,i}$  &  $z$     & $E_\text{iso}$ & $E_\text{p,i}$ & Type  & Experiment & Refs  & $EH$ Type & $EHD$ Type \\
        &    s    &      &  $10^{51}$ erg  & keV    &  &   &  	&   &    \\
 \hline

101213A	&	21.6	&	0.414	&	5.4	$ \pm $ 	0.4				&	354	$_{-	42	}^{+	58	}$	&	II	&	Konus-Wind & \citetalias{	tsv17}	& 1.80 II & 0.39 II \\
101219A	&	0.30	&	0.718	&	6.51	$ \pm $ 	0.36				&	1014	$_{-	96	}^{+	110	}$	&	I	&	Konus-Wind & \citetalias{	tsv17}	& 4.79 I & 8.75 I \\
101219B 	&	21.9	&	0.55 	&	5.9	$ \pm $ 	0.4				&	109	$ \pm $ 	12				&	II+SNsp	& Swift	 & \citetalias{	can17, qin13}	& 0.54 II & 0.12 II \\
101225A 	&	3790	&	0.847 	&	12	$ \pm $ 	3				&	70	$ \pm $ 	37				&	II+SNph	&	Swift & \citetalias{	can17}		& 0.26 II & 0.004 II \\
110205A	&	79.8	&	2.22 	&	560	$ \pm $ 	60				&	715	$ \pm $ 	239				&	II	& Swift	 & \citetalias{	qin13}		& 0.57 II & 0.06 II \\
110213A	&	11.9	&	1.46 &	69	$ \pm $ 	2				&	242	$_{-	17	}^{+	21	}$	&	II	&Swift	 & \citetalias{	qin13}		& 0.45 II & 0.13 II \\
\hline

 \end{tabular}
\end{table*}

\begin{table*}


 \contcaption{}  \label{ta1}

 \begin{tabular}{lccccccccc}
  \hline
 GRB    &  $T_\text{90,i}$  &  $z$     & $E_\text{iso}$ & $E_\text{p,i}$ & Type  & Experiment & Refs  & $EH$ Type & $EHD$ Type \\
        &    s    &      &  $10^{51}$ erg  & keV    &  &   &  	&   &    \\
 \hline

140508A	&	73.8	&	1.027 	 &	225	$ \pm $ 	15				&	446	$_{-	26	}^{+	28	}$	&	II	&	Konus-Wind & \citetalias{	tsv17}	& 0.51 II & 0.06 II \\
140512A	&	81.8	&	0.725 	&	72.5	$ \pm $ 	6.1				&	826	$_{-	136	}^{+	202	}$	&	II	&	Konus-Wind & \citetalias{	tsv17}	& 1.49 II & 0.17 II \\
140515A$^*$	&	3.2	&	6.32 	&	53.8	$ \pm $ 	5.8				&	376	$ \pm $ 	108				&	II	&	Swift & \citetalias{	gcn16284, gcn16269}	& 0.76 II & 0.43 II \\
140518A$^*$	&	10.6	&	4.707 	&	59.8	$ \pm $ 	5.8				&	251	$ \pm $ 	43				&	II	&	Swift & \citetalias{gcn16306, gcn16301}	& 0.49 II & 0.15 II \\
140606B 	&	4.9	&	0.384 	&	2.5	$ \pm $ 	0.2				&	352	$_{-	37	}^{+	46	}$	&	II+SNsp	& Konus-Wind	 & \citetalias{	can17,gcn16374}	& 2.44 II & 1.10 II \\
140619B	&	0.14	&	2.67	&	26.4	$ \pm $	1.2				&	4844	$ \pm $	1068				&	I	&	Fermi	& \citetalias{zha18a}	&	13.07 I &	35.42 I \\
\hline

 \end{tabular}
\end{table*}

\begin{table*}


 \contcaption{}  \label{ta1}

 \begin{tabular}{lccccccccc}
  \hline
 GRB    &  $T_\text{90,i}$  &  $z$     & $E_\text{iso}$ & $E_\text{p,i}$ & Type  & Experiment & Refs & $EH$ Type & $EHD$ Type \\
        &    s    &      &  $10^{51}$ erg  & keV    &  &   &  	&   &    \\
 \hline

170604A	&	12.9	&	1.329 	&	47	$ \pm $ 	6				&	512	$_{-	112	}^{+	168	}$	&	II	&	Konus-Wind & \citetalias{	gcn21247}	& 1.10 II & 0.31 II \\
170607A$^*$	&	14.5	&	0.557 	&	9.15	$ \pm $ 	0.05				&	226	$ \pm $ 	18				&	II	&	Fermi & \citetalias{	bha16, gcn21240}	& 0.93 II & 0.25 II \\
170705A$^*$	&	7.6	&	2.010 	&	180	$ \pm $ 	20				&	467	$_{-	66	}^{+	93	}$	&	II	&	Fermi & \citetalias{	bha16, gcn21298, gcn21299}	& 0.59 II & 0.21 II \\
170817A	&	0.50	&	0.00968 	&	(4.7	$ \pm $ 0.7) $\times$ 10$^{-5}$				&	65.6	$_{-	14.1	}^{+	35.3	}$	&	I	& Fermi & 	\citetalias{poz18a}	& 35.3 I & 50.0 I \\
170903A$^*$	&	13.6	&	0.886 	&	8.65	$ \pm $ 	0.06				&	180	$ \pm $ 	25				&	II	&	Fermi & \citetalias{	bha16, gcn21799}	& 0.76 II & 0.21 II \\
171010A 	&	116.7	&	0.3285 	&	180	$ \pm $ 	5.5				&	227.2	$_{-	8.0	}^{+	9.3	}$	&	II+SNsp	&	Konus-Wind & \citetalias{	gcn22003, gcn22096}	& 0.29 II & 0.03 II \\
\end{tabular}
\end{table*}

Table~\ref{ta1} contains observational parameters of the GRB sample.


\bsp	
\label{lastpage}
\end{document}